\documentclass[12pt,preprint]{aastex}

\usepackage{graphics,epsfig}
\usepackage{natbib}
\usepackage{lscape}

\def\spose#1{\hbox to 0pt{#1\hss}}
\def\lta{\mathrel{\spose{\lower 3pt\hbox{$\mathchar"218$}}
     \raise 2.0pt\hbox{$\mathchar"13C$}}}
\def\gta{\mathrel{\spose{\lower 3pt\hbox{$\mathchar"218$}}
     \raise 2.0pt\hbox{$\mathchar"13E$}}}

\def\figure#1#2 {\par{\narrower\noindent {\bf Fig. #1}
   \hskip 2mm #2\par}\bigskip\noindent}
\def\table#1#2 {\par{\narrower\noindent {\bf Tab. #1}
   \hskip 2mm #2\par}\bigskip\noindent}

\def\registered{{\ooalign{\hfil\raise .00ex\hbox{\scriptsize R}\hfil\crcr\mathhexbox20D}}}

\usepackage{amsmath}
\usepackage{accents}
\newlength{\dhatheight}

\slugcomment{Version: \enskip \today}

\shorttitle{Exocomets in the 47~UMa System}

\shortauthors{Cuntz, Loibnegger, \& Dvorak}

\begin{document}


\title{
Exocomets in the 47~UMa System: \\
Theoretical Simulations including Water Transport
}

\bigskip
\bigskip

\author{Manfred Cuntz$^1$, Birgit Loibnegger$^2$, Rudolf Dvorak$^2$}

\bigskip

\affil{$^1$Department of Physics, University of Texas at Arlington, \\
Arlington, TX 76019, USA}
\email{cuntz@uta.edu}

\bigskip

\affil{$^2$Institute of Astronomy, University of Vienna, T\"urkenschanzstr. 17 \\
A-1180 Vienna, Austria}
\email{birgit.loibnegger@univie.ac.at; rudolf.dvorak@univie.ac.at}


\begin{abstract}
Motivated by ongoing discoveries of features (most likely) attributable to
exo\-comets in various systems, this study examines the dynamics of possible
comets around 47~UMa.  Based on the assumption that most systems hosting planets
should also harbor leftovers from planet formation processes, comets are thus
also expected to exist in the system of 47~UMa.  This system is known to host
three Jupiter-type planets; however, based on stability analyses, additional
terrestrial planets in stable orbits might also be able to exist, including
planets in 47~UMa's habitable zone.  Furthermore, we also consider a possible
`Hilda'-planet.  The aim of our study is to explore the interaction of exocomets
with the Jupiter-type planets in the system and examine the probability of
cometary collisions with the planets, including possible Earth-mass planets
located in the habitable zone.  Moreover, we investigate the transport of
water onto the Earth-mass planets, including quantitative estimates.  It is
found that the Earth-mass planets would be able to receive some water, but
much less than currently present on Earth.  We also checked if the comets form families,
but no families were found.  Finally, the capture of comets in close orbits
and the possibility of small clouds formed when comets come too close to the
star and disintegrate are also part of our work. 
\end{abstract}


\clearpage


\section{Introduction}
\label{sec:intro}

Following the detection of absorption lines in the spectra of $\beta$ Pictoris
\citep{beu90} identified to vary on short time scales, it has been argued that
these lines are produced by clouds due to the evaporation of comets passing in
front of the star.  Additional results for $\beta$ Pictoris have been reported
by \cite{kie14}.  Evidence for comets in other exoplanetary systems, also
referred to as exocomets, has also been reported by, e.g., \cite{zuc12} and
\cite{wel13}.  Examples include 49~Ceti, a young A-type star, as well as,
e.g., HD~21620, HD~42111, HD~110411, HD~145964, and HD~183324.  Generally,
the evidence about possible exocomets is based on the interpretation of variable
Ca~II~K absorption line profiles.  The concept of exocomets was previously also
mentioned by \cite{ji05}, among others, from a dynamical viewpoint.
Recently, \cite{coo16} reviewed the realistic
detectability of close interstellar comets in the view of new methodologies, such
as the {\it Large Synoptic Survey Telescope} (LSST).  Note that besides exocomets,
the term `falling evaporating bodies' (FEBs) is sometimes used as well, especially
if a more general term is preferred.

Knowledge from the Solar System indicates that comets are small icy bodies ($\sim$0.1
to $\sim$100~km in diameter) that represent leftovers from planet formation \citep{bra08,ahe11}.
In fact, they may have played a pivotal role in the delivery of water and basic
organic substances from the outskirts of the Solar System to the terrestrial planets,
especially Earth and perhaps Mars as well; see, e.g., \cite{ray17} for recent theoretical
results.  Nevertheless, the efficiency of water transport by planetesimals, asteroids, and
comets is still part of previous and ongoing discussions \citep[e.g.,][]{gen08,ray17}.
The latter authors argue that the availability of comets to the inner stellar systems
is a natural consequence of the formation of giant planet(s); if correct, this result
would be highly relevant beyond the inherent physics and biochemistry of the Solar System. 

In the Solar System, cometary bodies have three known reservoirs, which are: the
Oort cloud, the Kuiper belt, and the Main belt \citep[e.g.,][]{hsi06}, which
originated from gravitational interactions of the planetesimals and asteroids with
the giant planets during the phase of planet formation \citep{bra08}. Assuming that
gravitational scattering is common during the stage of planet formation we conclude,
that there have to be reservoirs for small bodies in other planetary systems, too.
As mentioned above, features attributable to exocomets have been observed in the
spectra of many stars, with focus on A-type stars \citep[e.g.,][]{zuc12,wel13}.
Those systems are around mainly young stars ($\sim$5~Myr), which are supposedly
also accompanied by planet formation.  However, as pointed out by
\cite[e.g.,][and references therein]{coo16} comets are expected in other systems
as well, including interstellar space.  Very recently, \cite{rap18} reported the
first observations of (likely) transiting exocomets detected by {\it Kepler},
hosted by KIC~3542116 and KIC~11084727, both of spectral type F2~V.

The aim of the present work is to examine the dynamics of comets in the extrasolar
system of 47~UMa.  This system has some notable similarities to the Solar System,
including the spectral type and mass of the central star and the dominance of two
giant planets, which decisively impact the system's innermost parts.  Generally,
we want to study if the transport of water and prebiotic material could be possible
toward Earth-type planets, including planets located in 47~UMa's habitable zone (HZ).
In particular, material losses by comets will be calculated as well in order to
make predictions about the observability of comets in this type of system.  The
structure of the 47~UMa system, including the previous discovery of planets as well
as the extent of the stellar HZ, will be discussed in Sect.~2. The adopted methods
and the numerical setup are explained in Sect.~3.  Our results and discussion
are given in Sect.~4.  Section~5 conveys our summary and conclusions.


\section{The 47~UMa System}
\label{sec:system}

The system of 47~UMa has been regarded as showing significant similarities
to the Solar System\footnote{This statement was particularly genuine prior to
the discovery of the third Jupiter-type planet, 47~UMa~d.  However, 47~UMa~d is
located at a relatively large distance from the star, and thus is not expected
to significantly impact the inner part of the 47~UMa system, including
its HZ.}.  This assessment readily applies to the stellar spectral type,
effective temperature, metallicity, and mass; see Table~\ref{table1}.
In this study, we will use $T_{\rm eff} = 5890$~K, $M = 1.03~M_\odot$, and    
$R_{\ast} = 1.17~R_\odot$; these values have previously been determined by
\cite{hen97} and \cite{gon98}, with updates by \cite{kov03} and \cite{bel09}.

Previous studies have shown that 47~UMa harbors (at least) three Jupiter-type
planets in low-eccentricity orbits, designated as 47~UMa~b, 47~UMa~c, and
47~UMa~d, with semi-major axes of 2.1~au, 3.6~au, and 11.6~au (see Table~\ref{table2}).
Those detections have been made by \cite{but96}, \cite{fis02}, and \cite{gre10},
respectively, prominently based on the radial velocity method; 47~UMa~d was
identified using a Bayesian periodogram applied to the date.  Especially 
47~UMa~b and 47~UMa~c are found to be in almost circular orbits; the
planetary eccentricities are given as 0.032 and 0.098, respectively.
Thus, 47~UMa~b and 47~UMa~c are often regarded as Jupiter and Saturn analogs,
albeit that they are more massive and at closer proximity to their host star,
compared to Jupiter and Saturn to the Sun.  Scenarios for the formation of
the giant planets around 47~UMa have previously been described by \cite{the02}
and \cite{kor02}.

Estimates of 47~UMa's age are available as well.  Stellar ages can, for
instance, be obtained through analyzing the age-activity relations or 
isochrone fitting \citep{hen97}.  In case of 47~UMa, the age-activity
relationship of \cite{don93} implies 7 Gyr, in agreement with the estimate
of 6.9 Gyr by \cite{edv93} from isochrone fitting and work by \cite{ng98}.
More recent work by \cite{saf05} based on a variety of methods,
including those previously used by other authors, however, indicates
${\sim}6.03$~Gyr.  Therefore, we will adopt 6~Gyr in this study.  Thus,
47~UMa is considered a main-sequence star, although it already underwent
a noticeable evolutionary process, steadily approaching the subgiant stage.

Another important aspect pertains to the calculation of 47~UMa's
HZ\footnote{Previously, 47~UMa was studied by \cite{tur03} for its
potential regarding the {\it Search for Extraterrestrial Intelligence}
(SETI), and it received an overall positive assessment.  However, they
still decided not to include it in their {\it Catalog of Habitable Stellar
Systems} (HabCat), in part because 47~UMa's outer part of its HZ is
unavailable for hosting planets.  Nevertheless, 47~UMa continues to be
of great interest to the astrobiology research community.}.
\cite{kop13} provided detailed calculations for the limits of
stellar HZs using different criteria.  Based on this work, the inner and
outer limits for 47~UMa's general HZ (i.e., conservative estimates) are
given as 1.19~au and 2.04~au, respectively, whereas the limits for 47~UMa's
optimistic HZ are identified as 0.91~au and 2.13~au, respectively.
The latter values correspond to the climatological settings of recent
Venus / early Mars (RVEM) in the Solar System.  However, in case of
47~UMa, the outer part of the HZ is unavailable for hosting planets
owing to the influence of 47~UMa~b, the innermost system planet,
due to the initiation of orbital instabilities.  The truncation
occurs at about $\lta 1.6$~au, a number somewhat depending on the
values adopted as system parameters.  Detailed simulations on the truncation
of 47~UMa's HZ have been given by, e.g., \cite{nob02}, \cite{goz02},
and \cite{lau02}.  Previous theoretical studies about hypothetical
Earth-mass planets in the 47~UMa system, including analyses of orbital
stability, have been obtained by \cite{jon01}, \cite{cun03}, \cite{fra03},
\cite{asg04}, and \cite{ji05}.


\section{Methods and Numerical Setup}
\label{sec:methods}

\subsection{Comets}
\label{sec:methcomets}

As part of our study we assume that 47~UMa harbors an Oort-type cloud of comets
at the system's outskirt.  These comets are expected to be gravitationally
disturbed by, e.g., a passing star or through galactic tidal forces, which will
result in highly eccentric cometary trajectories allowing the comets to enter
the system's inner domain.  Cometary orbits and mechanisms of injections of
comets have previously been studied based by numerous authors
\cite[e.g.,][among others]{fou07,fou17,ric08,fen15}. 
The main goal of this study is to investigate how comets in the 47~UMa system
would be scattered or eventually captured by the system's planets.  In particular,
we are interested in exploring the probability of cometary collisions.
Specifically, we focus on cometary collisions with a fictitious planet assumed to
orbit within 47~UMa's HZ.  Therefore, we record the impact angles and
velocities of the comets' collisions, which will be used for further analyses,
including estimates of upper limits for the amount of water transported to the
planets. 

In our integrations we include the three known planets of 47~UMa
(see Table~\ref{table2}) as well as an additional (hypothetical) fourth planet
assuming three cases of initial semi-major axes:
\begin{itemize}
\item[] $a_1$ = 1 au
\item[] $a_2$ = 1.25 au
\item[] $a_3$ = 1.584 au~~~(`Hilda'-like orbit) 
\end{itemize}

We consider tens of thousands of fictitious comets assumed as massless\footnote{
Comets assumed as massless indicates that the star and the system planets (known or
hypothetical) are able to exert gravitational forces toward the comets but not
vice-versa.  Furthermore, the comets do not gravitationally interact with one another.}.
These comets have been evenly distributed in a sphere with initial conditions
given in Table~\ref{table5}.  Furthermore, an isotropic angular distribution
has been assumed for the cometary initial conditions owing to the lack of detailed
information about exocomets in that system.  As the comets have been assumed to
originate from an Oort Cloud analog, they have been placed in nearly hyperbolic
orbits; the integration time was set to 1~Myr.  Whenever a comet was ejected from
the system or underwent a collision, another comet was inserted with the same
`initial' conditions as the previous comet.  Since at the time of insertion the
configuration of the planets has changed, the newly inserted comet will exhibit
a different dynamics.

We also track the comets' semi-major axes and eccentricities to determine if
a comet was captured into a long or short periodic orbit.  Moreover, we will
inspect our results for analogs of cometary families as known for our
Solar System.  The integration of the equations of motion was performed using
Lie-transformation with adaptive step size control, which is particularly accurate
when modeling close encounters and collisions of celestial bodies
\citep[e.g.,][]{han84,egg10}.  The integrations were performed with a
highly variable step size.  The step size for the innermost planet is
at the beginning 10~days for a step-to-step precision of 10$^{-13}$.
However, due to the automatic step size control as part of the Lie-integrator,
it changes according to the involved masses and their mutual distances.
This means that at close encounters, i.e., just prior to a collision, the
step size attains values as small as a few hours.

\subsection{A `Hilda'-planet}
\label{sec:methhilda}

In the asteroid belt between Mars and Jupiter there exists an interesting
accumulation of asteroids residing in the region between 3.7~au and 4.2~au.
They orbit in a zone of mean-motion resonance (MMR) with Jupiter the one
in the orbital resonance 3:2.  The term {\it Hilda asteroids} --- nowadays
more than 1000 are known --- was coined after asteroid~153, Hilda,
was discovered by Johann Palisa\footnote{see {\tt{https://ssd.jpl.nasa.gov/sbdb.cgi?sstr=Hilda}} for details.}
in 1875.  Previous theoretical explorations of `Hilda' analogs in the 47~UMa
system have been given by \cite{lau02} and \cite{ji05}.

Because of the existence of 47~UMa~b, the innermost gas giant in the system
of 47~UMa with properties akin to Jupiter (albeit having a notably smaller
semi-major axis, i.e., 2.1~au versus 5.2~au; see Table~2), we also explore
the orbital properties of a hypothetical terrestrial planet assumed in the
region of the 3:2~MMR.  This planet would still be located in 47~UMa's HZ,
see \cite{kop13}, even though its outer segment (i.e., beyond ${\simeq}1.6$~au)
is unavailable due to the disturbances given by 47~UMa~b \citep[e.g.,][]{nob02,goz02}.
In the following, we investigate the interaction of the comets with this planet
as well, including the possible proliferation of water.


\section{Results}
\label{sec:results}

\subsection{Collisions}
\label{sec:collisions}

The aim of our study is to investigate the dynamics of (hypothetical) comets
in the system of 47~UMa, known to harbor three Jupiter-type planets with two
of them located closer to the host (i.e., at 2.1~au and 3.6~au, respectively)
compared to Jupiter to the Sun.  Additionally, we consider some fictitious
Earth-mass planets; see Sect.~\ref{sec:methcomets}.  Hence, we will evaluate
the statistics of collisions of comets with the planets (known or fictitious)
around 47~UMa.   Moreover, we will investigate the stability of the 3:2
resonance with the most massive planet in the system (47~UMa~b) that is in
analogy to the dynamics of the Hilda asteroids in the Solar System
(see Section~\ref{sec:hilda}).

A collision between a massless body (comet) and the Earth-mass planet
is considered an incident where the massless body comes as close to the
terrestrial planet as $r_{\rm Earth}$, taken as 6370~km.  At this distance,
the comet would hit the surface of the Earth-like planet.  Clearly,
the extent of the comet core itself can be assumed as negligible.

\begin{figure*} 
\centering
\begin{tabular}{c}
\epsfig{file=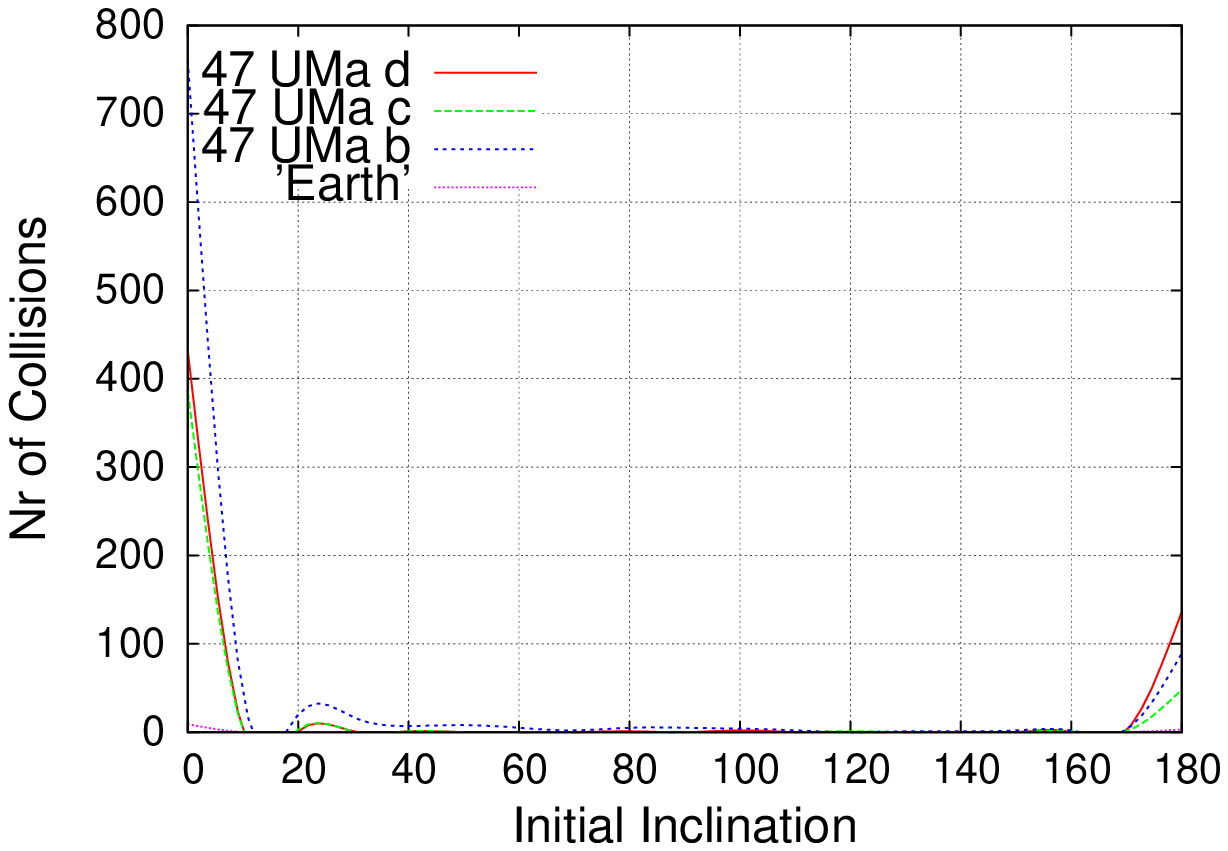,height=8cm} \\
\epsfig{file=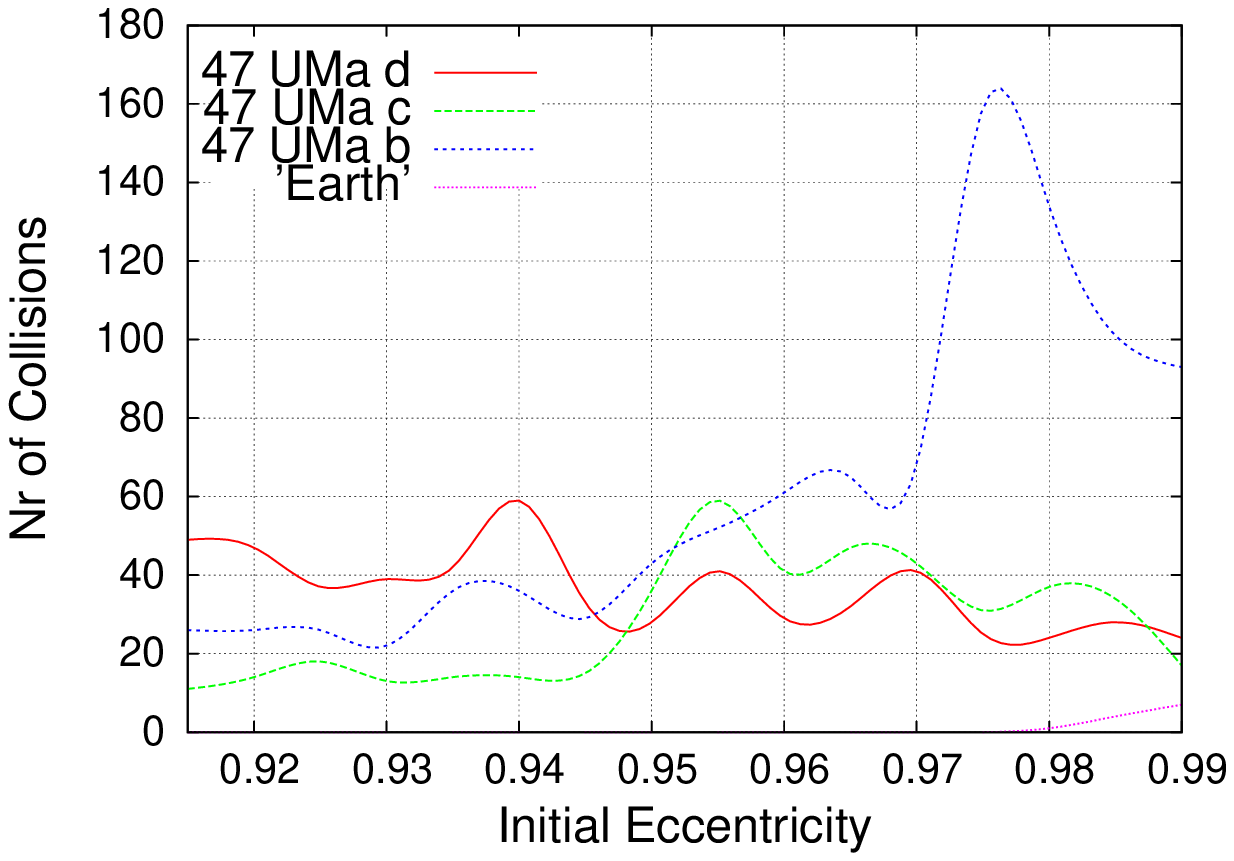,height=8cm}
\end{tabular}
\caption{Number of collisions with the three 47~UMa giant planets and the
fictitious Earth-mass planet at 1~au; furthermore, $a_{\rm ini,comet}= 80$~au
is assumed.  The most massive planet 47~UMa~b undergoes most collisions with
the comets.  Collision with comets of initially small inclinations
(i $<$ $20^{\circ}$) is most common for all planets (top panel).  Moreover,
the initial eccentricity of the comets has a significant influence on the
collision probability (bottom panel).  Comets with initially low eccentricity
($e < 0.945$) cannot reach the inner planetary system.  Thus, the number of
collisions for comets with these initial conditions is higher for the outermost
planet 47~UMa~d.  As soon as the initial eccentricity exceeds 0.945, comets can
reach the inner planetary system and thus experience close encounters with
47~UMa~b. The number of collisions peaks at 160 observed collisions with planet b
for an initial eccentricity of the comets 0.975.  For this value of initial
eccentricity, the comets have an initial perihel distance of 2.0~au, which is
slightly inside of the orbit of planet b.
}
\label{Fig1}
\end{figure*}

\begin{figure*} 
\centering
\begin{tabular}{c}
\epsfig{file=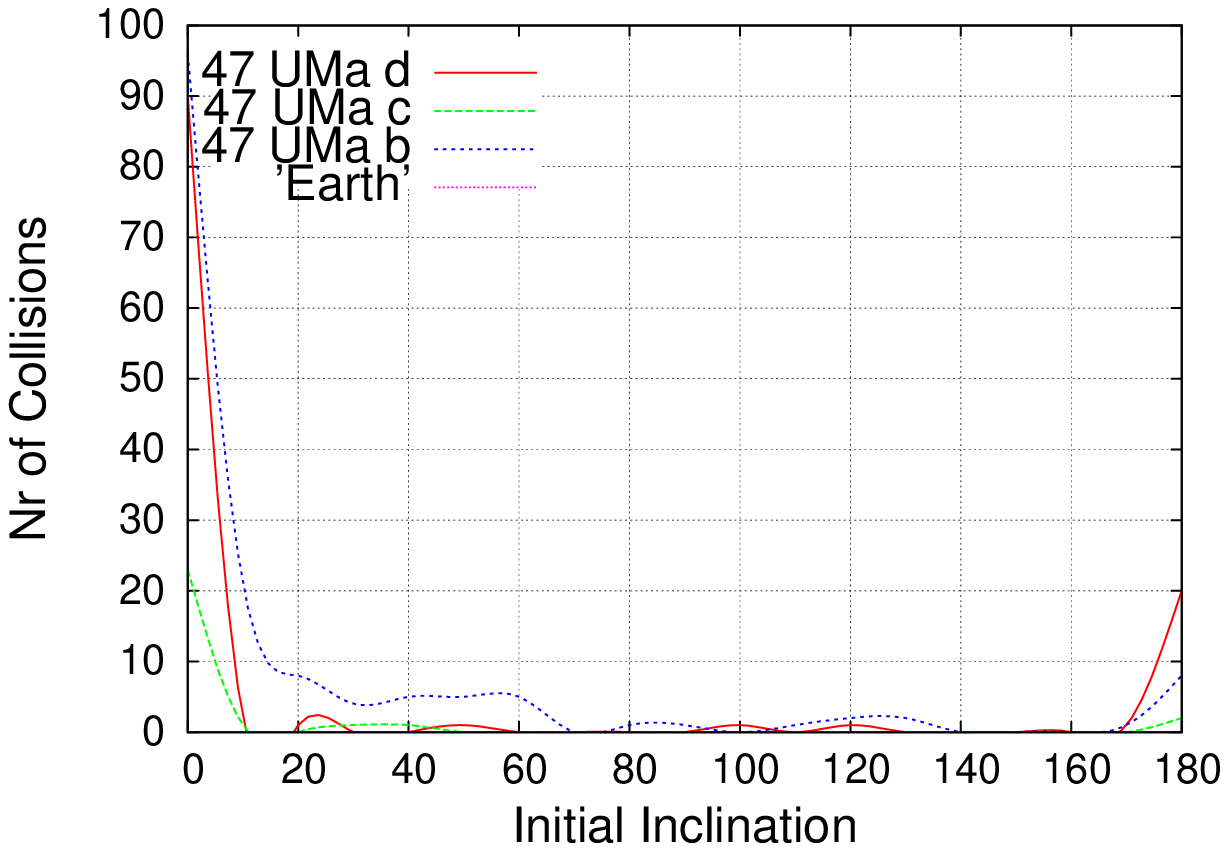,height=8cm} \\
\epsfig{file=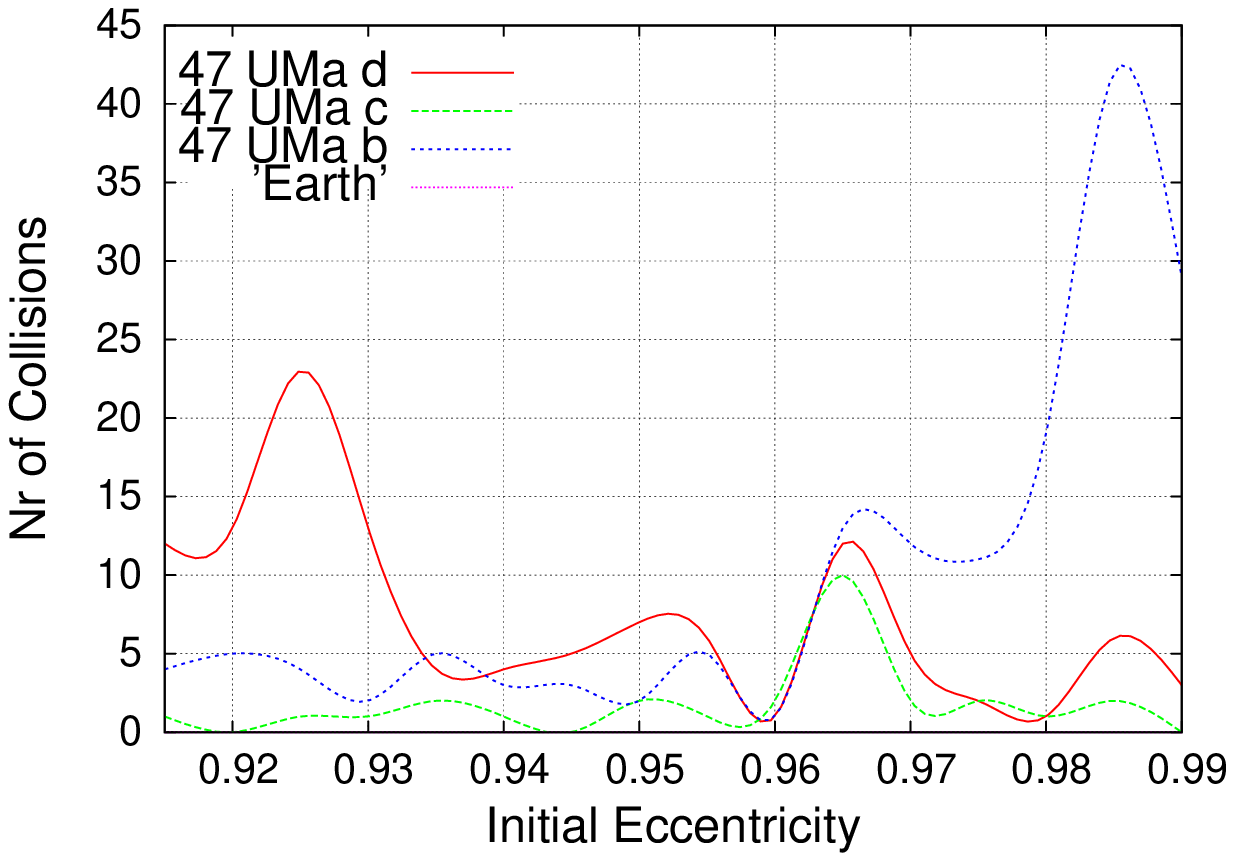,height=8cm}
\end{tabular}
\caption{
Same as Fig.~\ref{Fig1}, however the comets are started with an
initial semi-major axis of $a_{\rm ini,comet}= 120$~au.  The peak of
collisions for comets with initial eccentricities of 0.925 for the outermost
planet 47~UMa~d shows that now a large number of comets is only able to reach
the far-out orbit of this planet, which encounters the majority of collisions.
}
\label{Fig2}
\end{figure*}

A key motivation for our study of cometary collisions is that we want to
explore the possible water transport by comets to the Earth-mass planets.
In Figure~\ref{Fig1} we show the number of collisions of comets with the
four planets included in our integrations.  Clearly, the probability of
a collision strongly depends on the initial conditions of the comets
(i.e., initial inclination and initial eccentricity).  Comets with
relatively low initial eccentricities are unable to reach the inner
planetary system.  Consequently, they cannot collide with any of the
Earth-mass planets put close to or within the HZ unless they are scattered
inward by the 47~UMa system planets.  In fact, the most massive planet
in the system, 47~UMa~b, experiences the most collisions.  Furthermore, it
is more likely for comets of low initial inclinations to collide with any
of the planets.  This behavior is found for all initial configurations; it
is also visible in Figure~\ref{Fig2}.  It occurs as a result from the choice
of initial parameters for the planets orbiting 47~UMa, which in lack of
knowledge about their inclinations have been placed in coplanar orbits.  

\begin{figure*} 
\centering
\begin{tabular}{c}
\epsfig{file=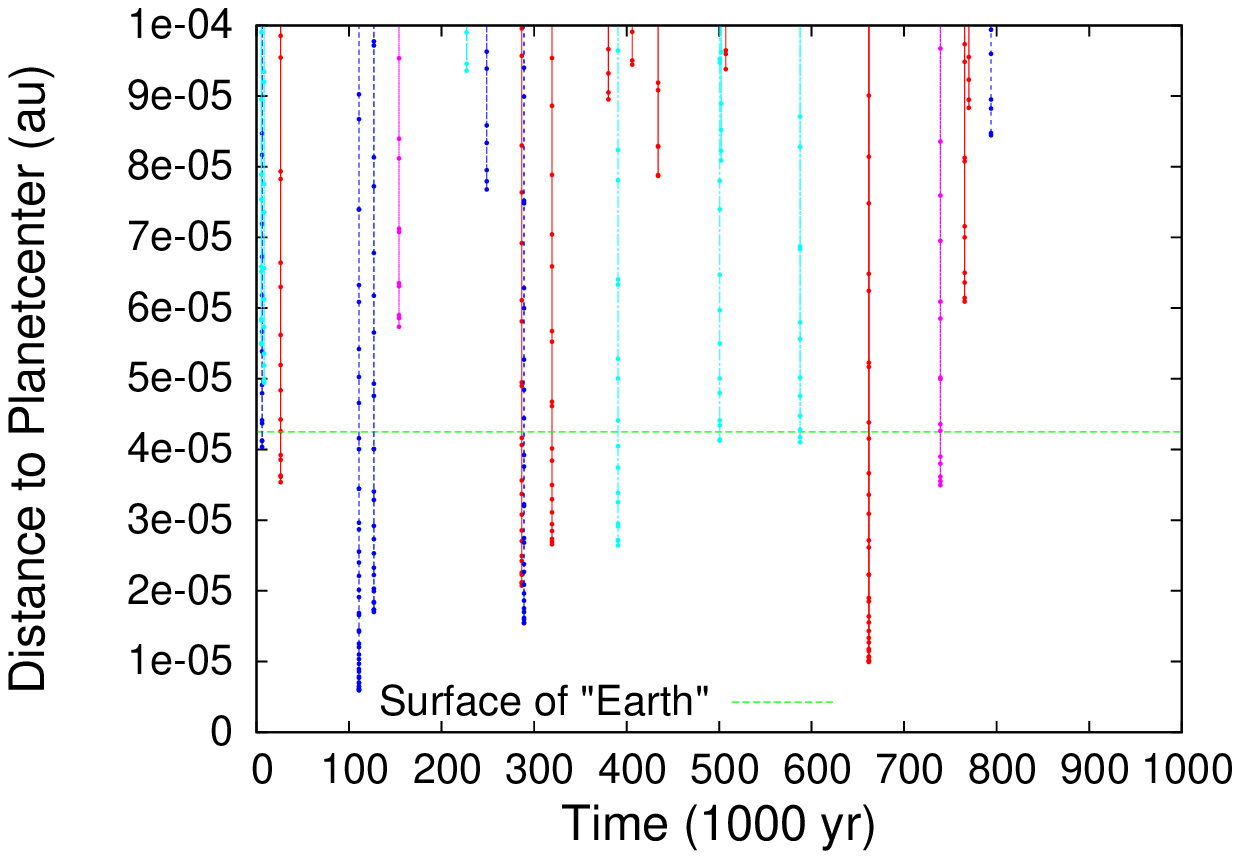,height=6cm} \\
\epsfig{file=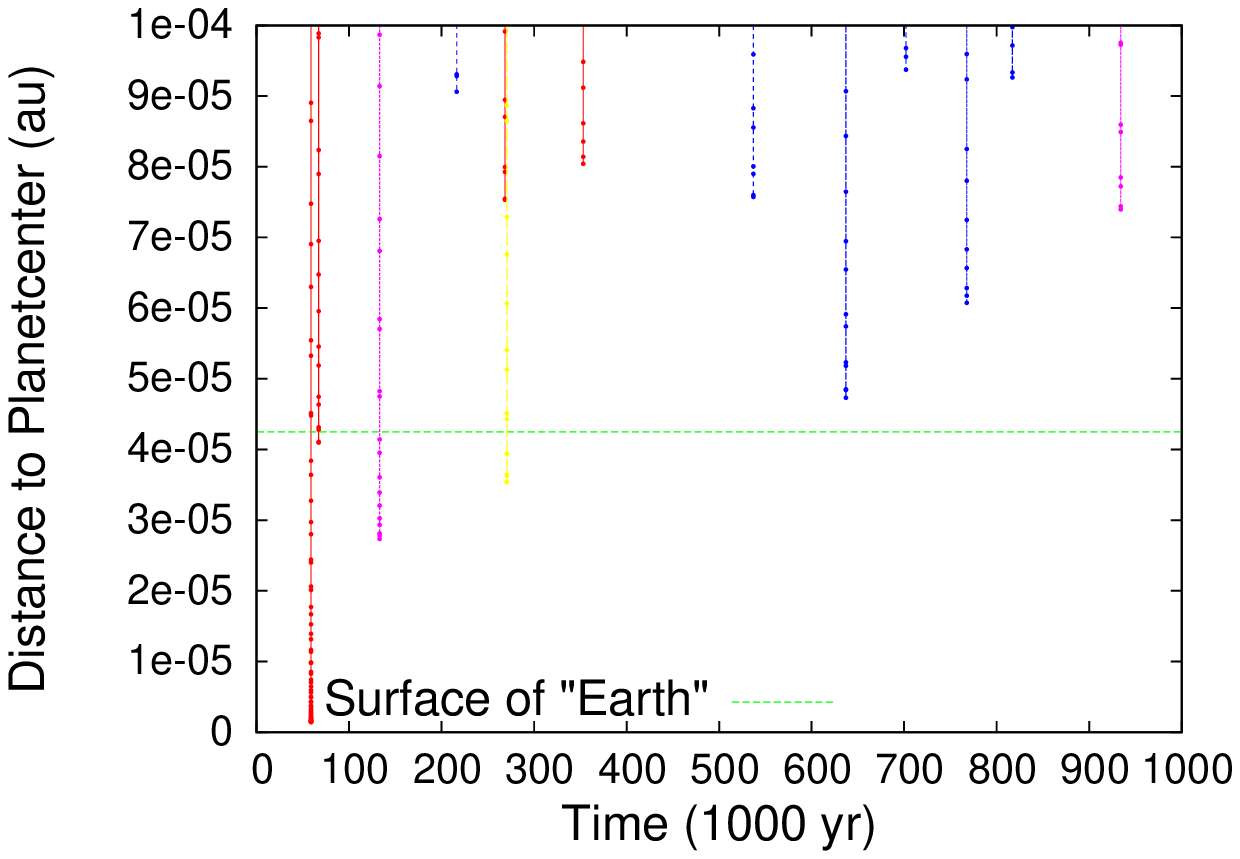,height=6cm} \\
\epsfig{file=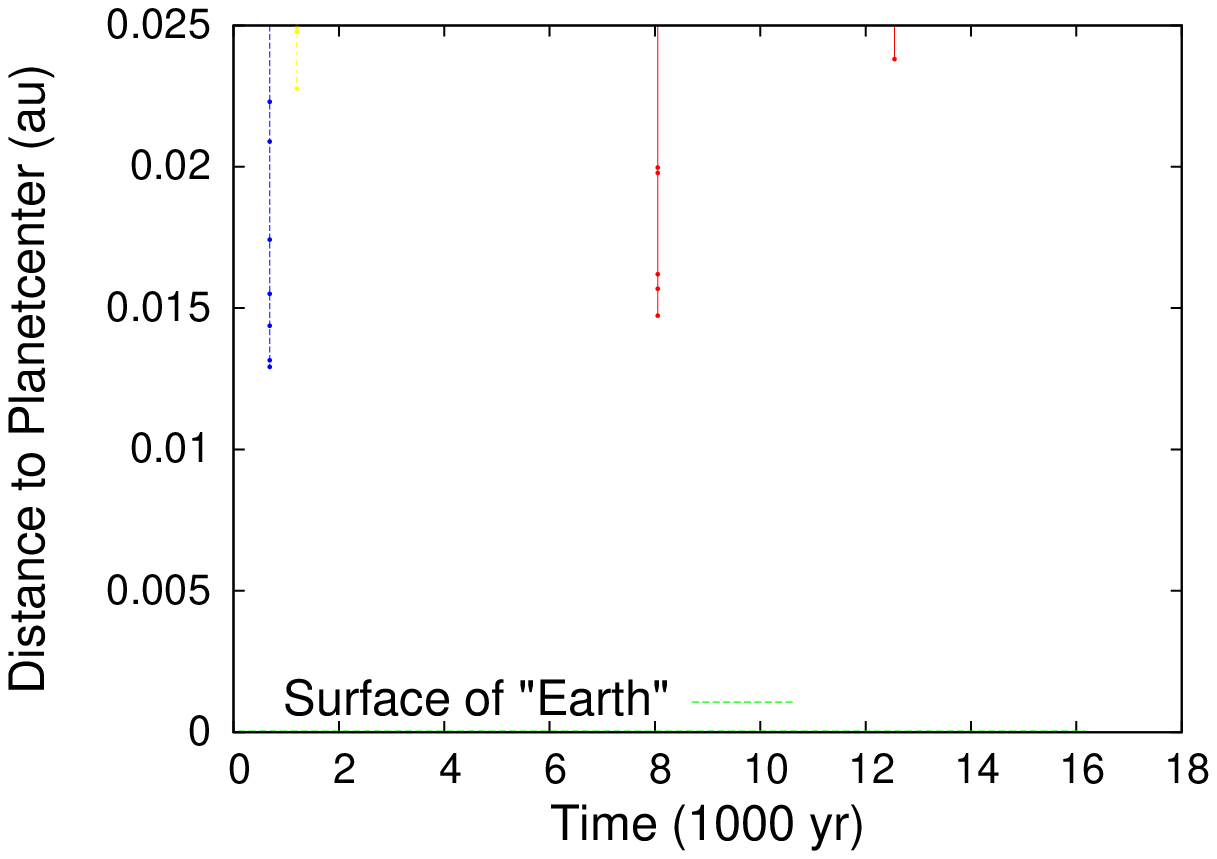,height=6cm}
\end{tabular}
\caption{
The three plots convey the real collisions with the Earth-mass planet
at different locations, which are: 1~au (top panel), 1.25~au (middle panel),
and 1.584~au, the `Hilda'-planet (bottom panel).  Collisions with the comets,
treated as massless test particles, already occur during the first 1~Myr of
integration.  The green lines mark the surface of the planet.  Note that
all lines crossing the green line denote real collisions.  It is found
that the number of collisions decreases as the Earth-mass planet is put
closer to the orbit of the most massive planet, 47~UMa~b.  Note the
different ranges for the $y$-axes.  Regarding the `Hilda'-planet, no real
collisions with the planet occur this close to 47~UMa~b.
}
\label{Fig3}
\end{figure*}

In Figure~\ref{Fig3}, we depict cometary collisions with the Earth-mass planet
at different locations.  The surface of the planet is marked with a green line.
One can see that comets cross this line, indicating that collisions have occurred.
During these collisions transport of water and organic material is expected to
take place.  Interestingly, the closer the Earth-mass planet is moved toward the
most massive planet (47~UMa~b), the lesser number of cometary collisions occur.
The Earth-mass planet is placed first at 1 au, and at this orbit it suffers
from most collisions.  Putting it in an orbit closer to 47~UMa~b considerably
reduces the number of collisions (see Figure~\ref{Fig3}).

When we put the Earth-mass planet at 1.584 au (an orbit in 3:2~MMR resonance
with 47~UMa~b analog to the Hilda family of asteroids in the Solar System),
no collisions are found to occur at all, which makes water transport to a
Hilda analog through cometary collisions in 47~UMa impossible.  We conclude
that the proximity to the most massive planet in the system and the occurrence
of orbital resonances prevent Earth-mass planets in the distance range-as-evaluated
from being hit by comets; all comets are scattered away by 47~UMa~b.  This behavior
indicates that (1) Earth-mass planets in the 47~UMa system in resonance with the
most massive planet can be orbitally stable, and (2) those planets will be spared
from collisions with comets or other objects entering the system on high eccentric
trajectories owing to the gravitational influence and scattering caused by 47~UMa~b.
Thus, the biggest planetary object, partly due to the virtue of resonances, acts
like a shield, protecting the inner planetary system.   


\subsection{Water transport}
\label{sec:watertransport}

In order to estimate an upper limit for water transport by comets, we assume
that a comet consists by 20\% of water \citep{ful17}.  For simplicity, we
also assume that all the water transported by the comet is transferred to
the fictitious terrestrial planet in case of a collision.  This is only a
first step in the framework of our envisioned studies, and we are aware that
the collision geometry will probably reduce the amount of transferred water
by a significant extent.  The collision details such as impact angles and
impact velocities are stored from our calculations; in fact, more detailed
collision studies are anticipated.

We made the following assumptions for the terrestrial planet orbiting at 1~au:
A comet has an estimated mass of $10^{14}$~kg, which is approximately the
the average mass of comets measured in the Solar System.  In the beginning
we put 100 comets on eccentric trajectories about the star.  They amount to a
total mass of $10^{16}$~kg, which entails an estimated water content
for the total cometary population of $2 \times 10^{15}$~kg.  If a comet in engulfed by
the star or is ejected from the system, it is replaced by a new comet. This
can add up to a very high number of comets intruding the planetary system
during our integrations.  Earth's oceans consist of $1.5 \times 10^{21}$~kg
of water.  This means that in our initial comet population, 0.00013\% of Earth's
oceans are incorporated.
This amount of water, for example, would also correspond to 8.8\% of the combined
body of water\footnote{The total amount of water for the Great Lakes has been
estimated as 5439 cubic miles.  Information as reported by the
U.~S. Environmental Protection Agency; see
{\tt http://www.epa.gov/glnpo/physfacts.html}} of the Great Lakes, located on
the Canada -- United States border.

For example, we calculated the transported mass of water for comets with
initial semi-major axis of 80~au and no inclination:
118,557 comets have been induced in addition to the initially 100 comets
of the system.  This amounts to a total number of 118,657 comets integrated
for 1 Myr in total, with an estimated total mass of $1.2 \times 10^{19}$~kg
and a total water content of $2.4 \times 10^{18}$~kg (expressed in Earth's
oceans: $0.16\%$).  Of these 118,557 comets, 9 collided with the Earth-mass
planet depositing $1.8 \times 10^{14}$~kg of water, which corresponds to
0.000012\% of Earth's oceans or 0.8\% of the Great Lakes.
This amount of water is delivered during an integration time of 1~Myr.

If this value is extrapolated to the age of 47~UMa, given as 6 Gyr
(see Table~\ref{table1}), a total amount of water of $1.1 \times 10^{18}$~kg
(i.e., 0.073\% of Earth's oceans or 48.5 times the water of the Great Lakes)
is derived, thus able to reach the surface of
the terrestrial planet at 1~au.  For the total number of comets considered
with initial semi-major axes of 80~au, the following values are obtained:
2,174,536 comets plus 1900 initial ones amount to 2,176,436 comets.
In this combined number of comets, a total mass of water given as
$4.4 \times 10^{19}$ kg is included.
For this configuration we observed 12 collisions.  If perfect merging
is assumed, and the delivery of the total amount of water from the comets
to the Earth-mass planet (orbiting at 1~au in this scenario) occurs as
said, $2.4 \times 10^{14}$~kg of water are delivered to the planet.
This corresponds to 0.000016\% of Earth's oceans or 1.1\% of the water
of the Great Lakes, transfered to the planet during 1~Myr.  Extrapolating
this value to an age of 6~Gyr results in
$1.4 \times 10^{18}$~kg of water, equivalent to 0.093\% of
Earth's oceans or 61.7 times of the water contained in the Great Lakes.

Taking all the comets with initial semi-major axes between
$80~{\rm au} < a_{\rm ini, comet} < 200~{\rm au}$ into account, only 12 collisions
with the Earth-mass planet are observed at 1~au.  Comets
with higher initial semi-major axes are found not to be able to reach
the inner planetary system and collide with the terrestrial planet.  These
calculations have been repeated for the Earth-mass planet at 1.25 au and the
`Hilda'-planet.  The outcomes are given in Table~\ref{table2}.  We found that
the `Hilda'-planet does not receive any water because it does not encounter
any collisions with the comets.  This convincingly demonstrates the shielding
effect of the most massive planet, 47~UMa~b.  Additionally, the 3:2 resonance
--- where the `Hilda'-planet resides --- protects that planet from collisions
with the cometary bodies.

Since perfect merging has been assumed, i.e., the transfer of the total amount
of water of a comet to the Earth-mass planet, only an upper limit of water
reaching the surface of the Earth-type planet inserted in the system of 47~UMa
is described.  For the 47~UMa system, we realize that the number of collisions
with the small planet orbiting inside the three known system planets is very small.
Especially the most massive planet, 47~UMa~b, orbiting at 2.1~au from the star,
scatters most of the comets away, thus preventing them from entering the outer HZ
at about 1.6~au, the approximate distance of the `Hilda'-planet.  Water transport has
previously been estimated by \cite{ban15,ban17}, who considered water transport
via asteroids to planets in binary star systems.


\subsection{Captured comets}
\label{sec:capturedcomets}

Motivated by the findings of two distinct families of comets in the
$\beta$ Pictoris system \citep[see][]{kie14}, we also tried to distinguish
between different cometary families in the 47~UMa system among the comets
scattered into orbits of low semi-major axes and eccentricities; these comets
were considered captured.  Unfortunately, it was not possible to determine
different families of comets in this system.  Nevertheless, the capture of
comets into orbits was noted in our simulations due to gravitational
scattering by the massive planets of the system, especially 47~UMa~b. 

\begin{figure*} 
\centering
\begin{tabular}{c}
\epsfig{file=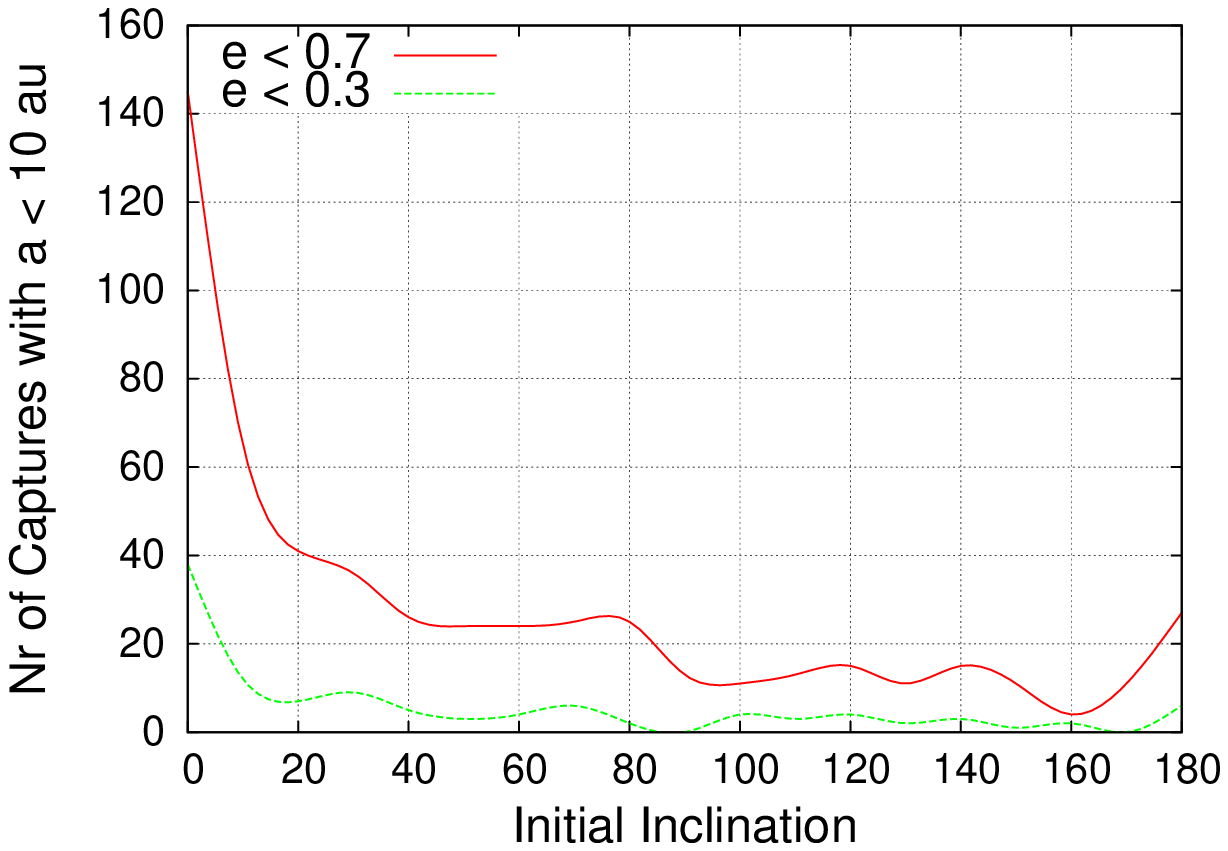,height=8cm} \\
\epsfig{file=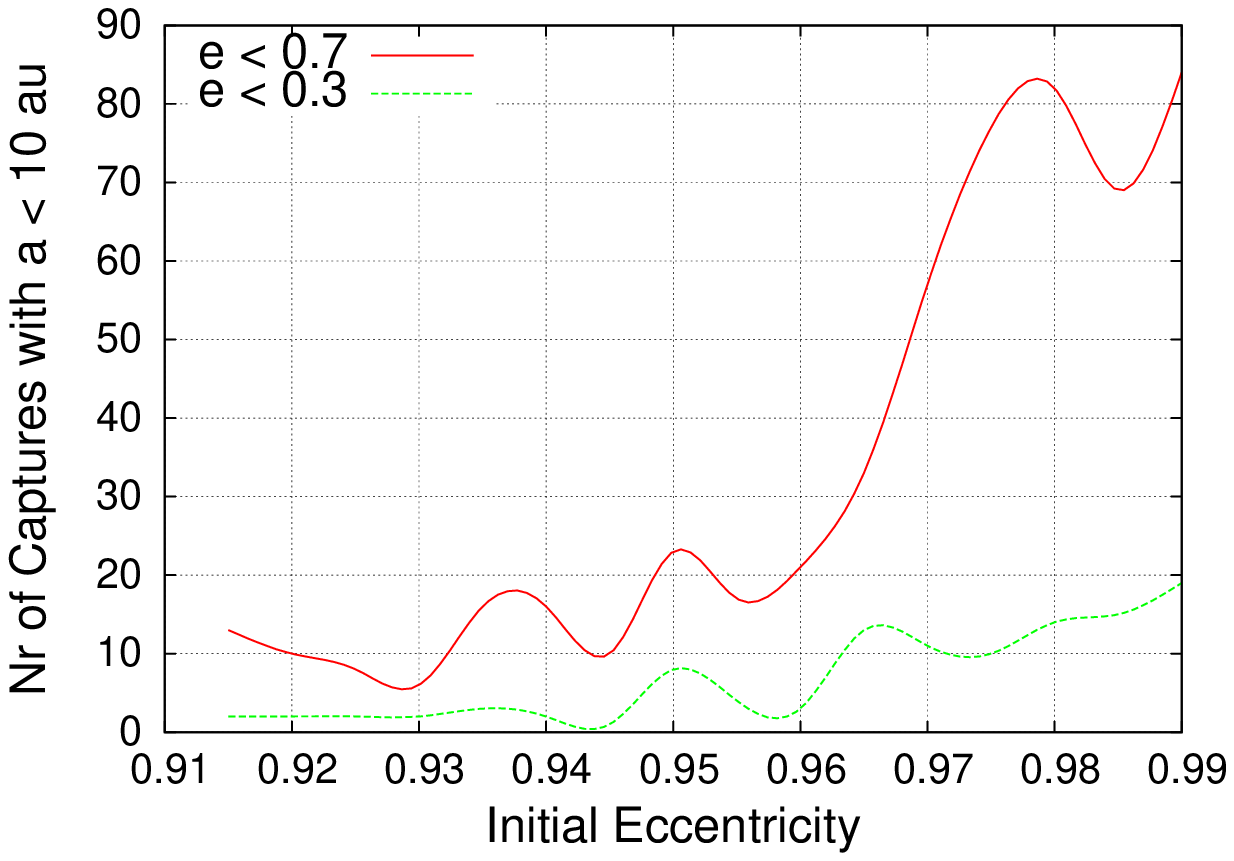,height=8cm}
\end{tabular}
\caption{
Number of captures of comets in orbits with $a<10$~au and, respectively,
$e<0.7$ and $e<0.3$ after an integration time of 1~Myr.  Results are
given for different values of initial inclination (top) and initial
eccentricity (bottom).  Note that captures occur for comets of all
types of initial conditions.  Nevertheless, the number of comets
scattered to an orbit with the desired values is highest for comets
with small initial inclinations and high initial eccentricities.
Only comets with high eccentricity are able to reach the inner parts
of the system and interact with the most massive planet 47~UMa~b, the
main perturber capable of scattering the small bodies into orbits with
low semi-major axis and eccentricity.
}
\label{Fig4}
\end{figure*}

Our integrations showed that the capture of comets into orbits with a
low semi-major axis and eccentricity is possible.  This can result in
comets of short-periodic orbits, maybe forming some kinds of
family analogs to our Solar System.  The lower the eccentricity of
the intruding comet, the higher is the probability to be scattered to
a less eccentric orbit (see Figure~\ref{Fig4}). Comets with initially higher
eccentricities are able to penetrate deeper into the planetary system and
reach the orbit of the most massive planet 47~UMa~b at a distance of 2.1~au.
Due to its mass this planet has the biggest influence on orbits of comets
entering the inner part of the 47~UMa system. 

If we also add the Earth-mass planet to the system, the big picture doesn't
change.  The reason is that this hypothetical planet, located closer than
47~UMa~b to the star, has barely any influence on the cometary scattering
process.  Interestingly, the number of captured comets decreases for the
Earth-mass planet placed at 1.25 as well as 1.584~au, relative to the case
for the Earth-mass planet at 1.0~au.


\subsection{Encounters with the planets}
\label{sec:encounterswiththeplanets}

Encounters of the comets with the planets are important for facilitating
scattering processes.  A close encounter is defined as the flyby of a comet
within 5 Hill radii of the planet.  A close encounter leads to a change of
the comet's trajectory and thus may lead to a capture or collision with one
of the planets, or the comet's ejection from the system.  When assumed
that comets in the proximity of the star start to lose mass by degassing,
it can thus also be assumed that material may be transferred from the comet
to the planet in case of a close encounter.

\begin{figure*} 
\centering
\begin{tabular}{c}
\epsfig{file=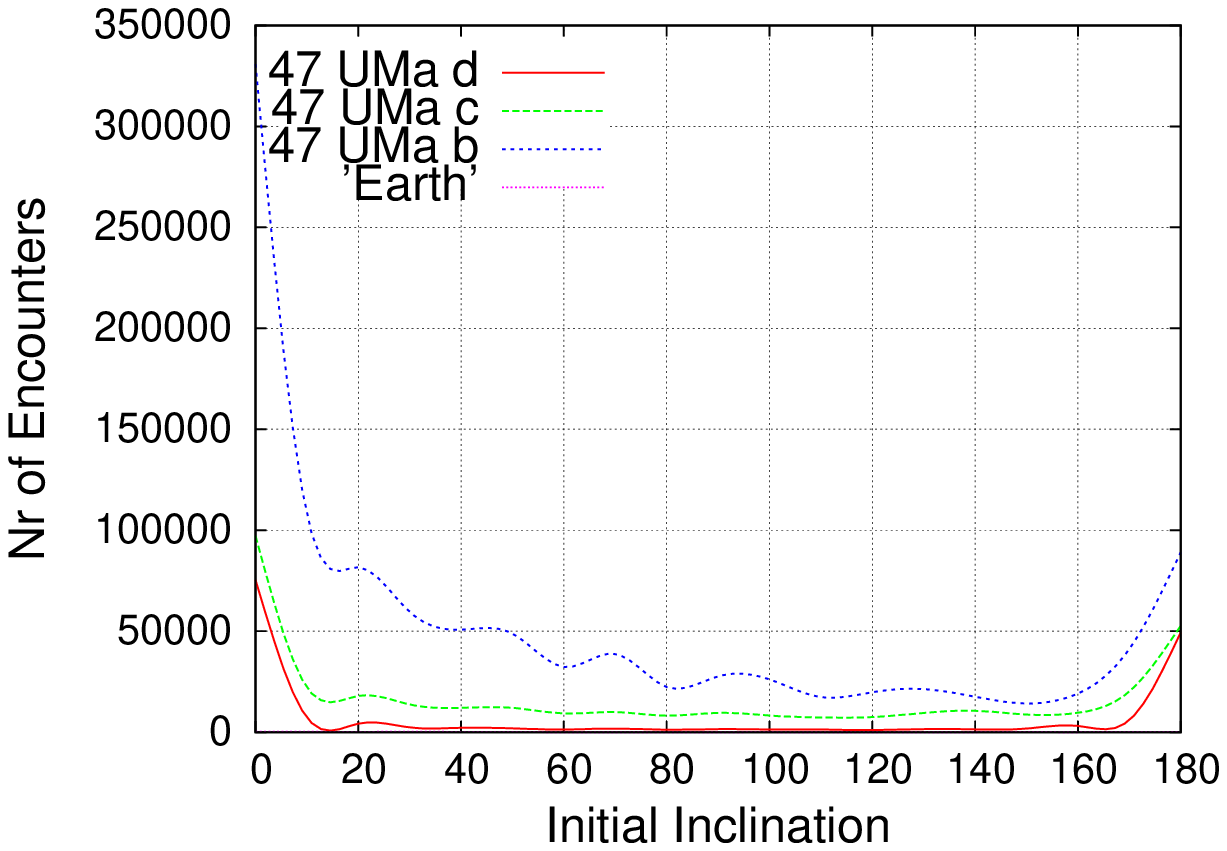,height=8cm} \\
\epsfig{file=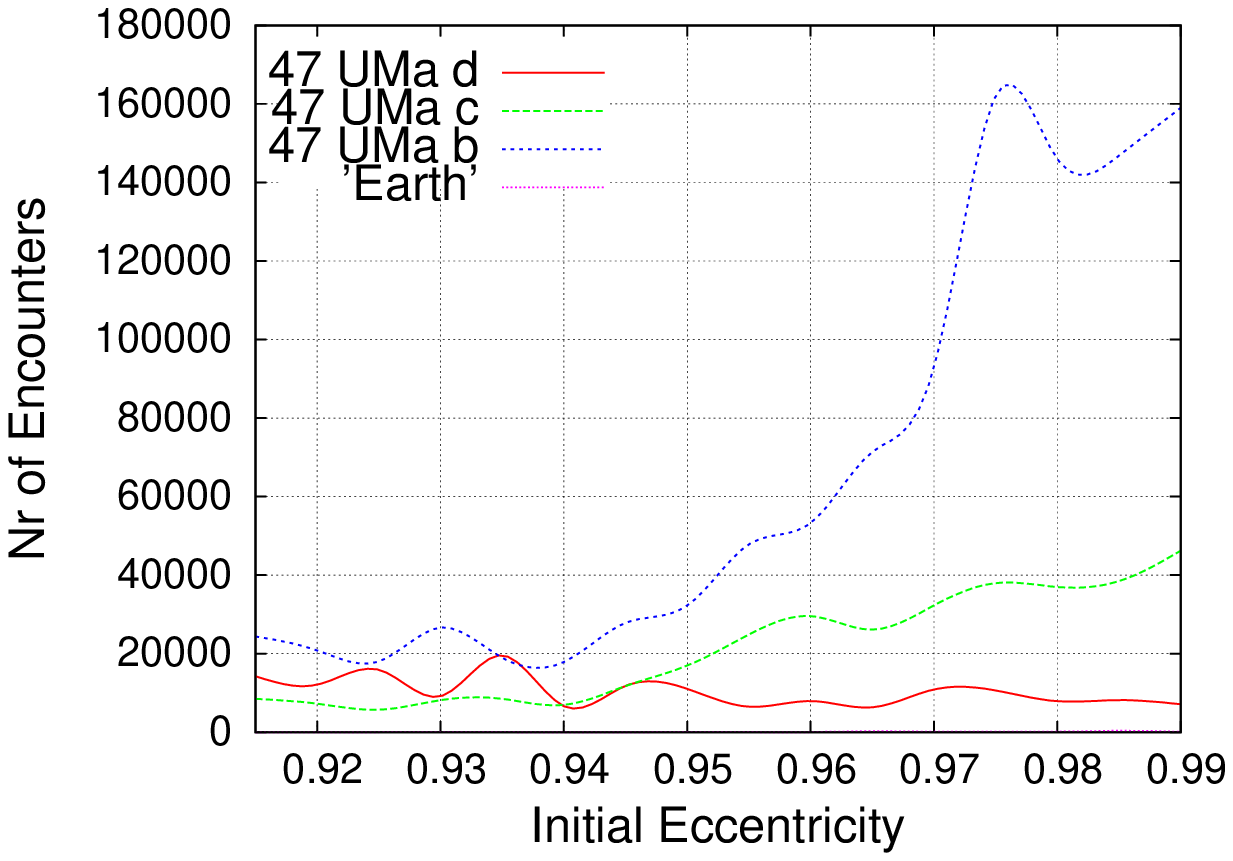,height=8cm}
\end{tabular}
\caption{
Number of encounters of the Earth-mass planet with the various 47~UMa
giant planets after it was put in a resonant orbit with 47~UMa~b, the
most massive planet of the system.  Note that the encounters of the
Earth-mass planet with the giant planets are very rare.  Planet 47~UMa~b,
the most massive planet, experiences most encounters, and thus has the
biggest influence on cometary scattering.
}
\label{Fig5}
\end{figure*}

The latter will lead to the transfer of cometary material onto the planet,
including, e.g., meteorite showers.  In Figure~\ref{Fig5} we show the number
of close encounters with the 4 planets in the 47~UMa system (i.e., the
3 known planets and a hypothetical `Earth').  The most massive planet,
47~UMa~b, experiences the most close encounters.  This result depends on
the initial conditions of the comets, as expected.  Comets with an initially
low eccentricity will not come as close to the star, largely because of
the presence of 47~UMa~b.  Thus, the number of encounters is remarkably
smaller than for comets with an initially high eccentricity.  As already
mentioned, due to the system's structure, the interaction between the
planets and comets with initially high inclinations is sparse, which is
the reason for the relatively low number of close encounters between
those objects. 


\subsection{Orbital distribution of comets}
\label{sec:orbitaldistribution}

\begin{figure*} 
\centering
\begin{tabular}{c}
\epsfig{file=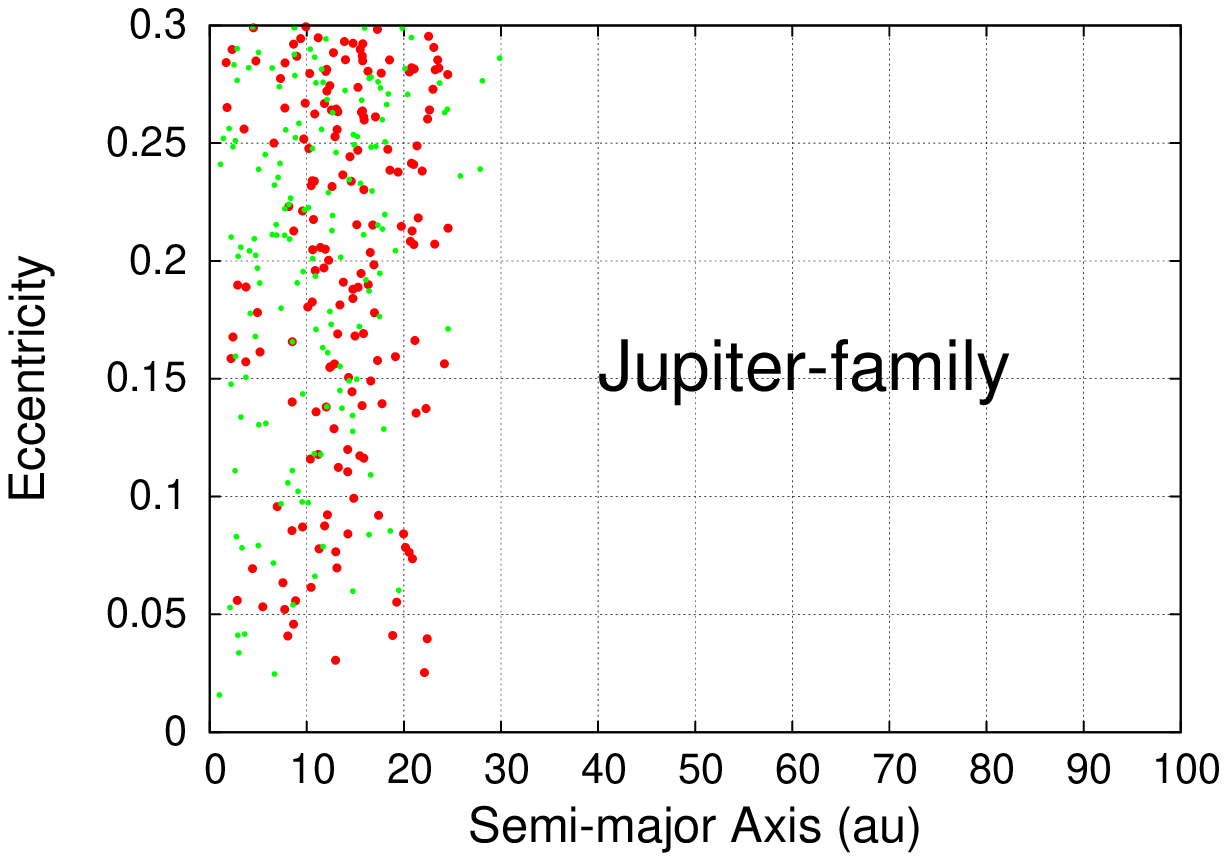,height=6cm} \\
\epsfig{file=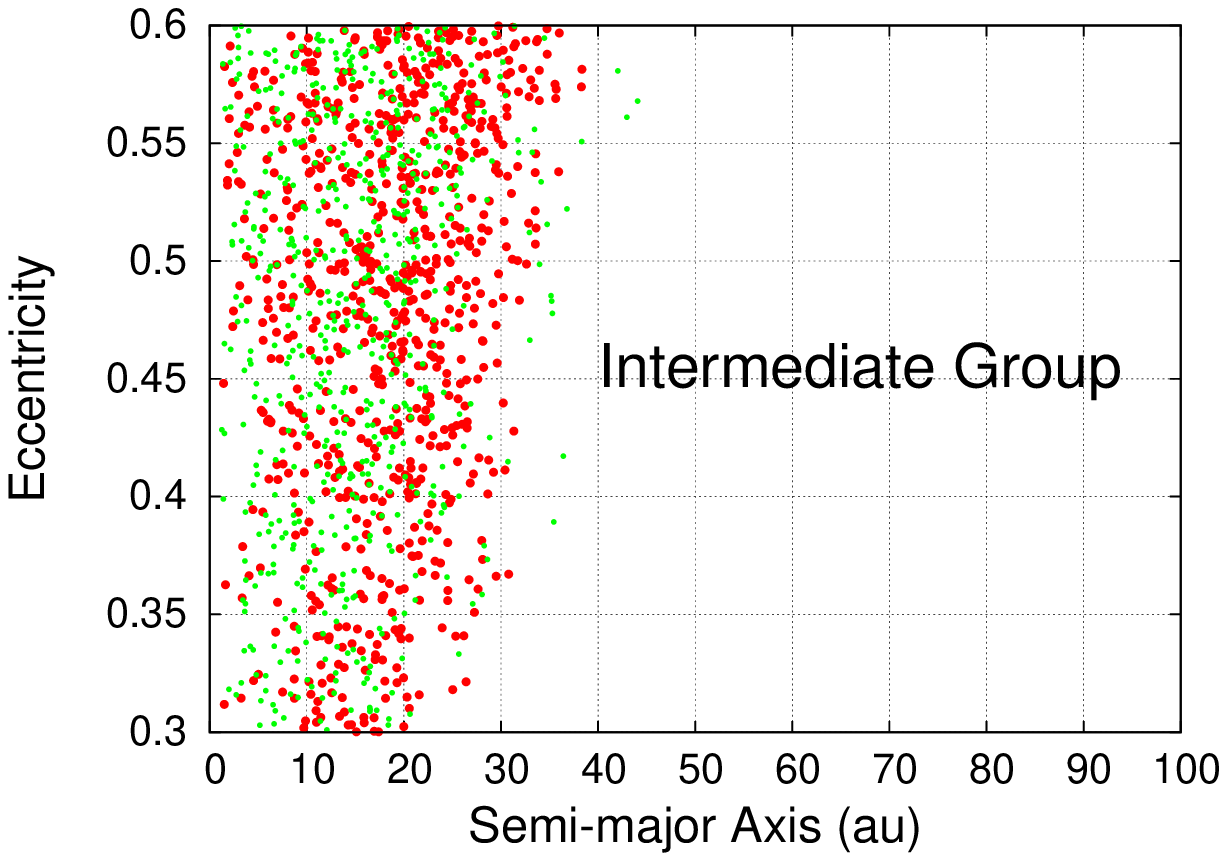,height=6cm} \\
\epsfig{file=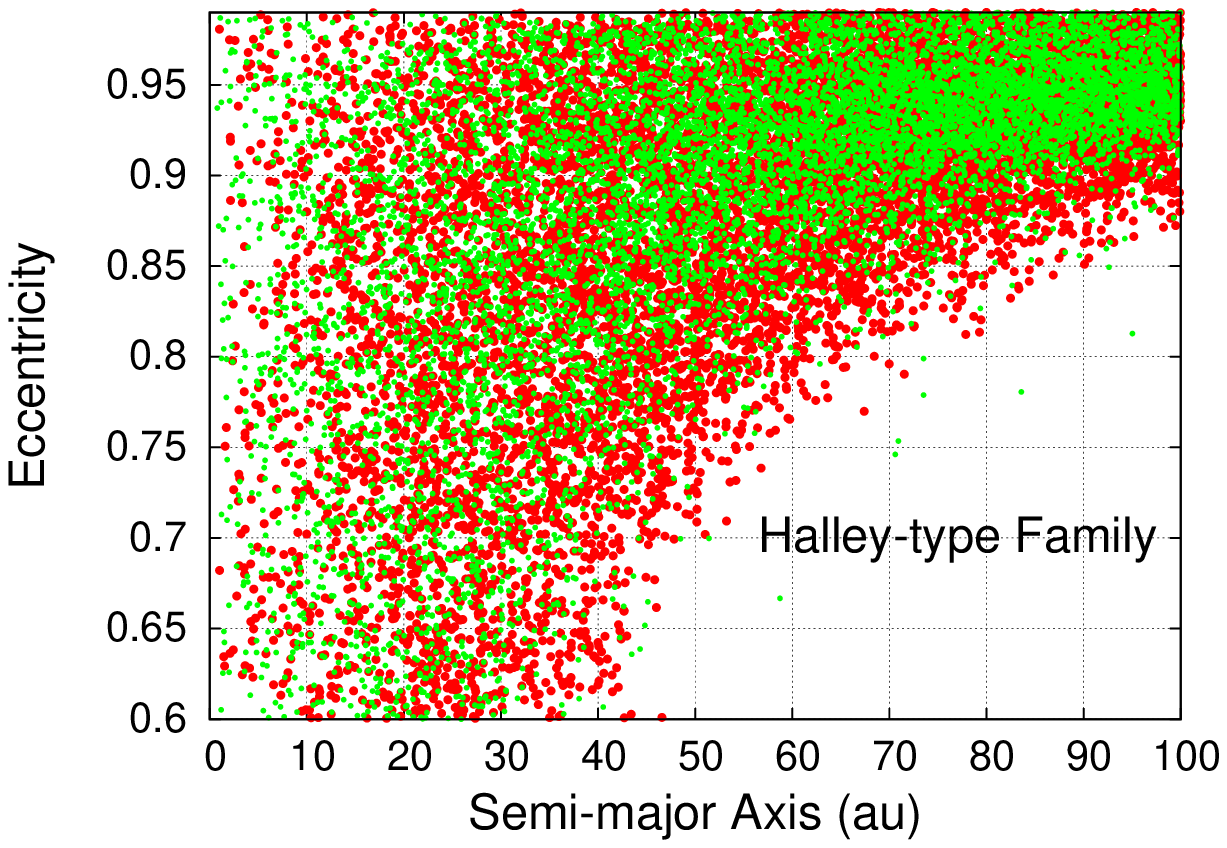,height=6cm}
\end{tabular}
\caption{
Distribution of the orbital elements semi-major axis and eccentricity
for the comets after an integration time of 1~Myr.  The green and red dots
correspond to comets ending up on prograde and retrograde orbits, respectively.
The distribution is close to uniform.  Both prograde and retrograde comets
are found in orbits with relatively small eccentricities and semi-major axes.
}
\label{Fig6}
\end{figure*}

The distribution of orbital elements of the trajectories (i.e., semi-major axis
versus eccentricity) identified for the comets after an integration time of
1~Myr is shown in Figure~\ref{Fig6}. The integrated system included an Earth-mass
planet originally placed at 1~au.  Furthermore, the comets started with an initial
semi-major axis of 80~au.  Regarding the initial distribution, half of the comets
were started in retrograde orbits relative to the 47~UMa planetary system.  It is
noteworthy that at the end of the integration time, half of the comet population
remained in retrograde orbits despite the shorter interaction time with the planets
due to their trajectories.

Our simulations did not reveal different families of comets in the system of
47~UMa; however, we are inclined to conclude that the system might contain
different comet populations.  There is a small population of comets being
scattered to orbits with very low eccentricities and small semi-major axes,
which might be analogous to the Jupiter-family comets of the Solar System
(see Figure \ref{Fig6}, left panel).  An even bigger population is found with
eccentricities between 0.3 and 0.6, but still small semi-major axes with values
$< 40$~au (Figure~\ref{Fig6}, middle panel). This intermediate population of comets
appears to be a bridge of the Jupiter-family-type comets to the Halley-type comets
revealed by the results of our calculations.  The latter yields the biggest number of
comets identified.  They have relatively high eccentricities ($e > 0.6$) and their
semi-major axes are mostly larger than 50~au (Figure \ref{Fig6}, right panel). 


\subsection{The `Hilda'-Planet}
\label{sec:hilda}

We also explored the influence of a planet in a Hilda-like orbit
(see ref~\ref{sec:methhilda}), originally placed at 1.584~au.  In order
to identify the stability regions for that additional planet, we checked
the entire region of the 3:2~MMR assuming a cloud of fictitious comets
as before.  In terms of their initial conditions, have been equally
distributed from $1^{\circ}$ to $360^{\circ}$ in the difference of the
perihelion asteroids -- gas giant, i.e.,
$\Delta (\omega_{\rm giant} - \omega_{\rm planet})$.  In varying the
semi-major axes, the eccentricities, and the inclinations of the comets,
we were able to identify the regions of greatest stability.

\begin{figure*} 
\centering
\begin{tabular}{c}
\epsfig{file=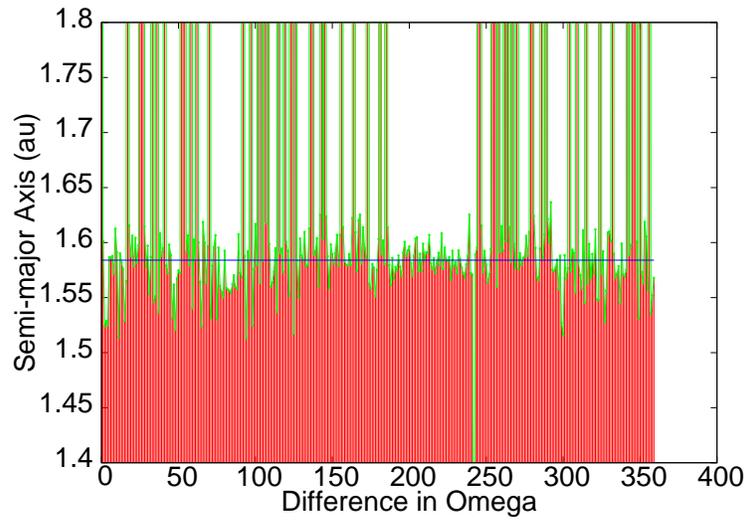,angle=270,width=10cm}
\end{tabular}
\caption{
Attained semi-major axes ($y$-axis) 
versus the differences in the perihelion of the gas giant and
the planet, $\Delta (\omega_{\rm giant} - \omega_{\rm planet})$
($x$-axis); the latter are equally distributed between $1^{\circ}$
and $360^{\circ}$.  The results refer to 360 fictitious massless
`Hilda'-planets after an integration time of 1~Myr.
Two stable windows are identified.
}
\label{Fig7}
\end{figure*}

\begin{figure*} 
\centering
\begin{tabular}{c}
\epsfig{file=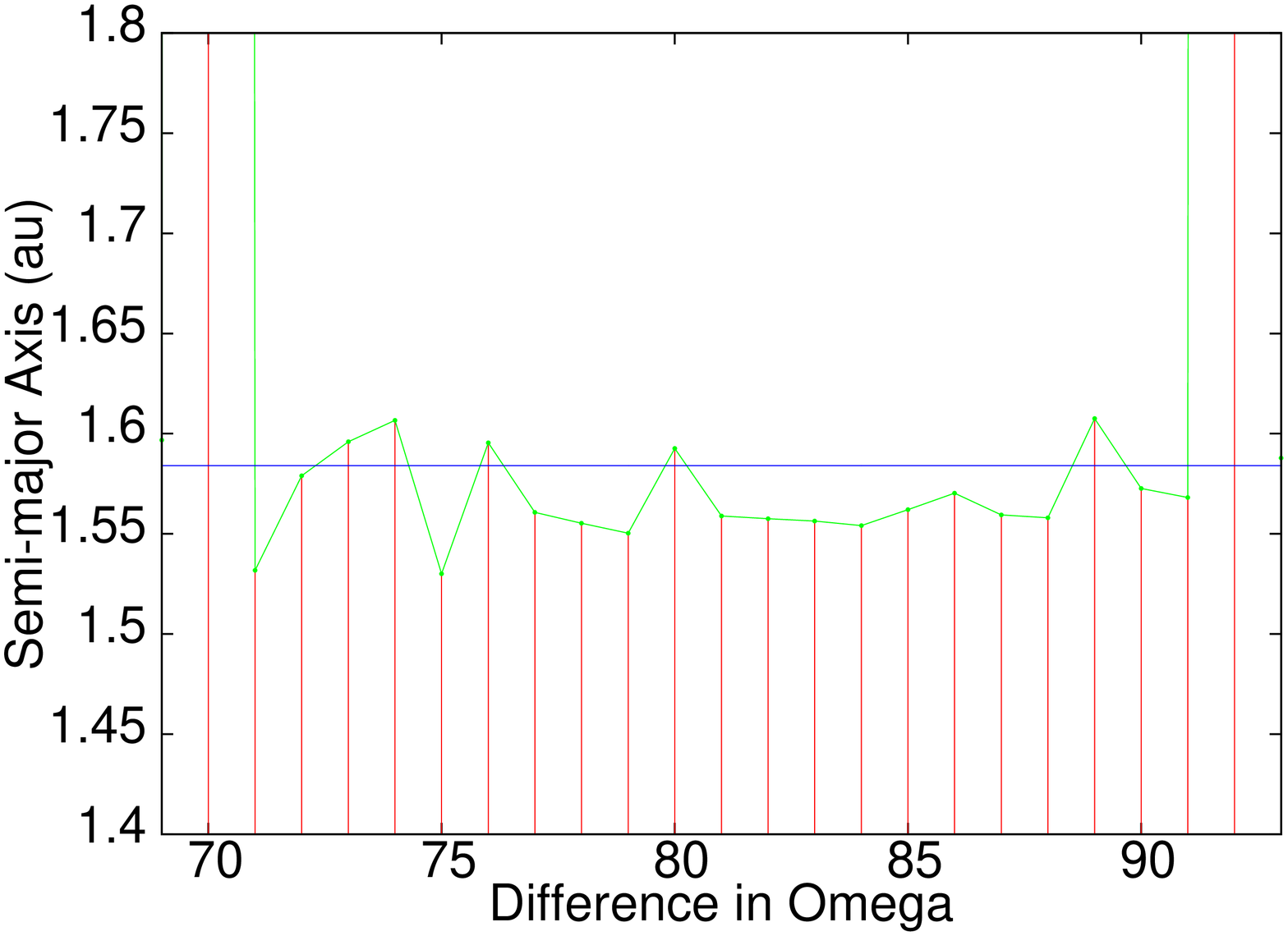,angle=270,width=10cm} \\
\epsfig{file=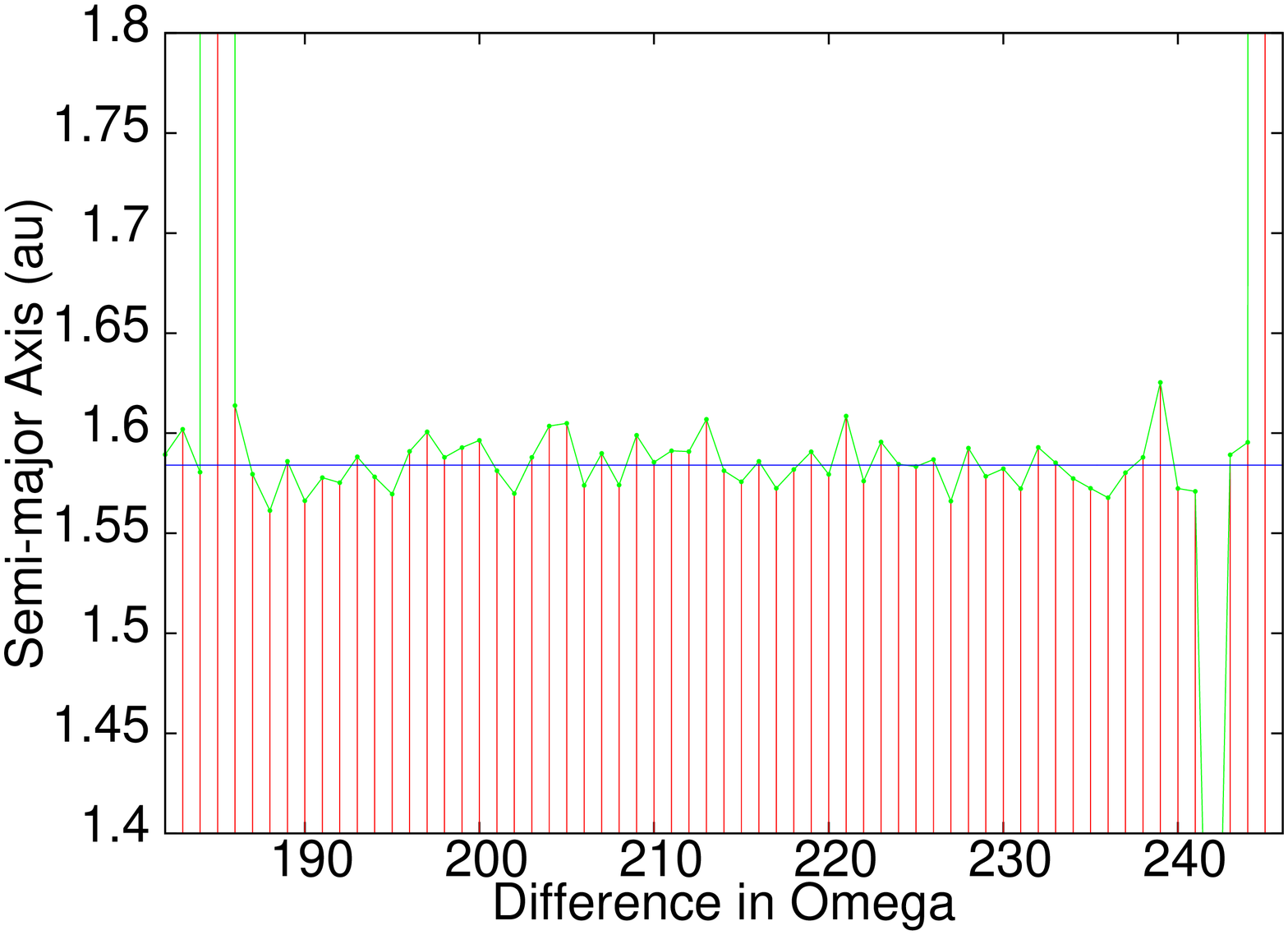,angle=270,width=10cm}
\end{tabular}
\caption{
Zoomed-in results of Fig.~\ref{Fig7} for a small window around $80^{\circ}$
(top) and a larger window around $215^{\circ}$ (bottom).
Axes as in Fig.~\ref{Fig7}.
}
\label{Fig8}
\end{figure*}

Figure~\ref{Fig7} depicts one of the diagrams semi-major axes versus
the differences in the perihelion longitude between the gas giant and
the fictitious comets.  From orbital consideration of the 3:2~MMR, we
expect two windows: the first around $85^{\circ}$ and the second around
$215^{\circ}$.  In fact, we can see from the plot that in our numerical
integrations two main stable windows appear about those values.  Between
the windows, some stable orbits for `Hilda'-planets can be seen; however,
these objects are expected to escape after longer times of integration.
A zoom of the two windows of Figure~\ref{Fig7} is depicted in Figure~\ref{Fig8}.
  
\begin{figure*} 
\centering
\begin{tabular}{cc}
\epsfig{file=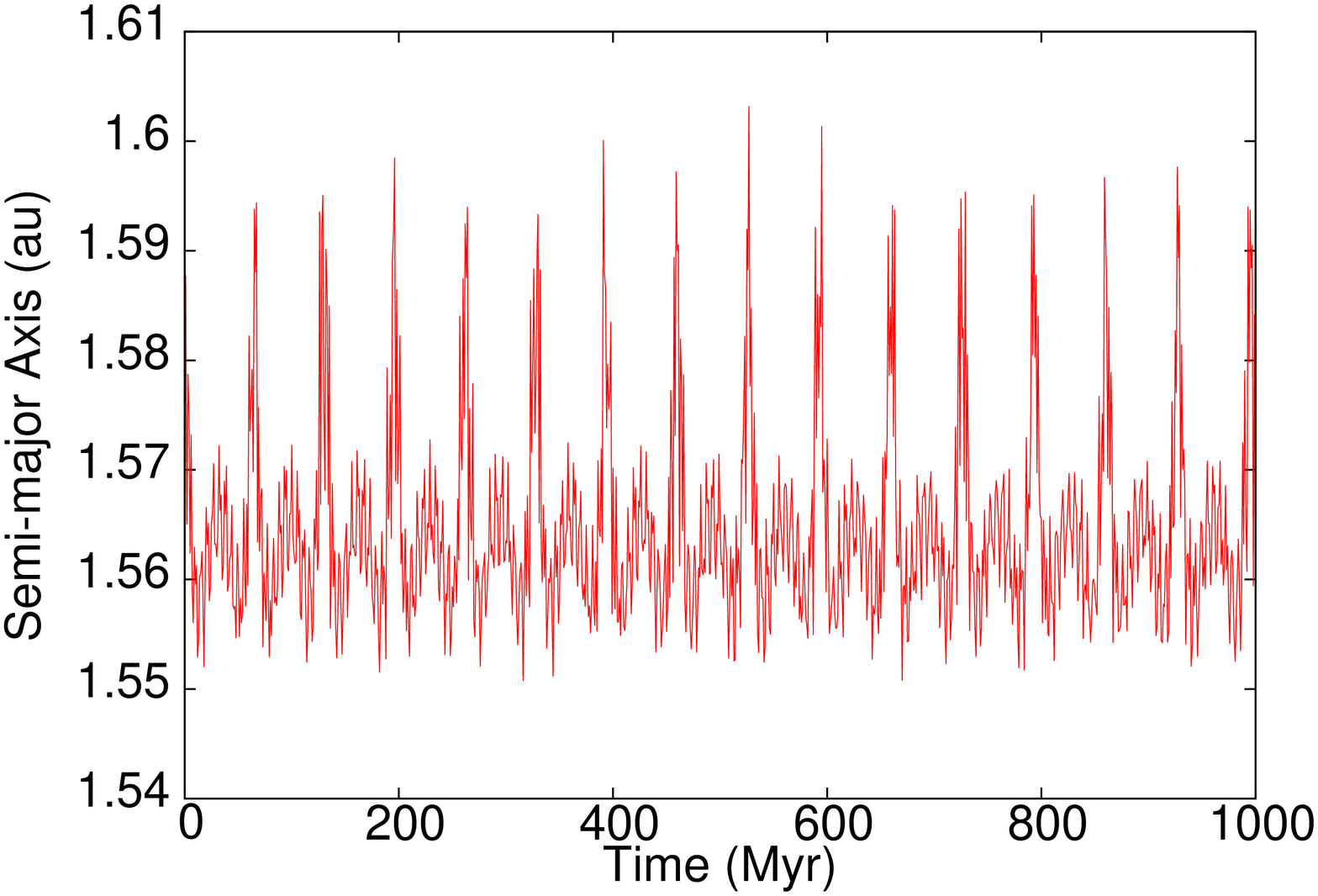,angle=270,width=8cm}
\epsfig{file=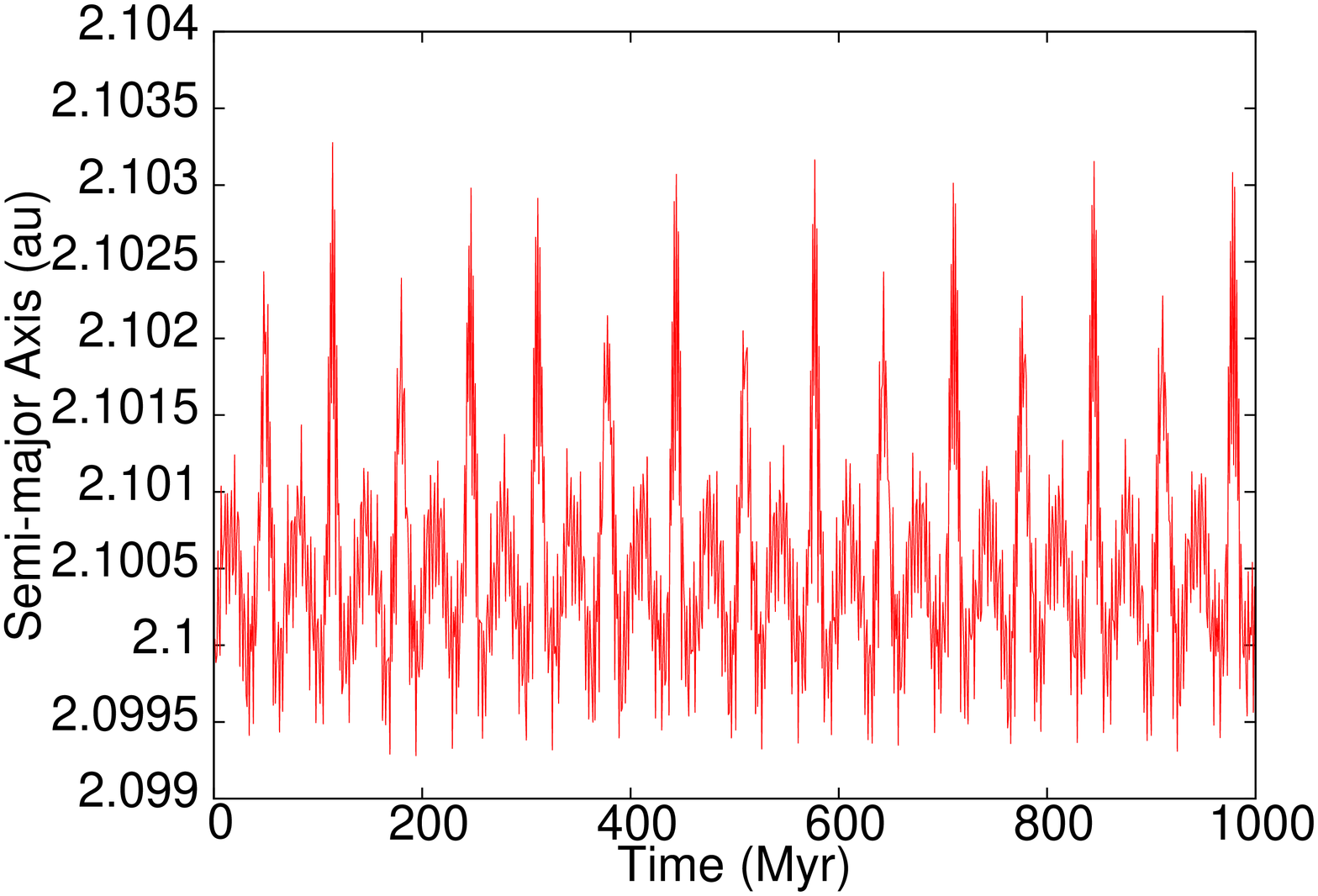,angle=270,width=8cm} \\
\epsfig{file=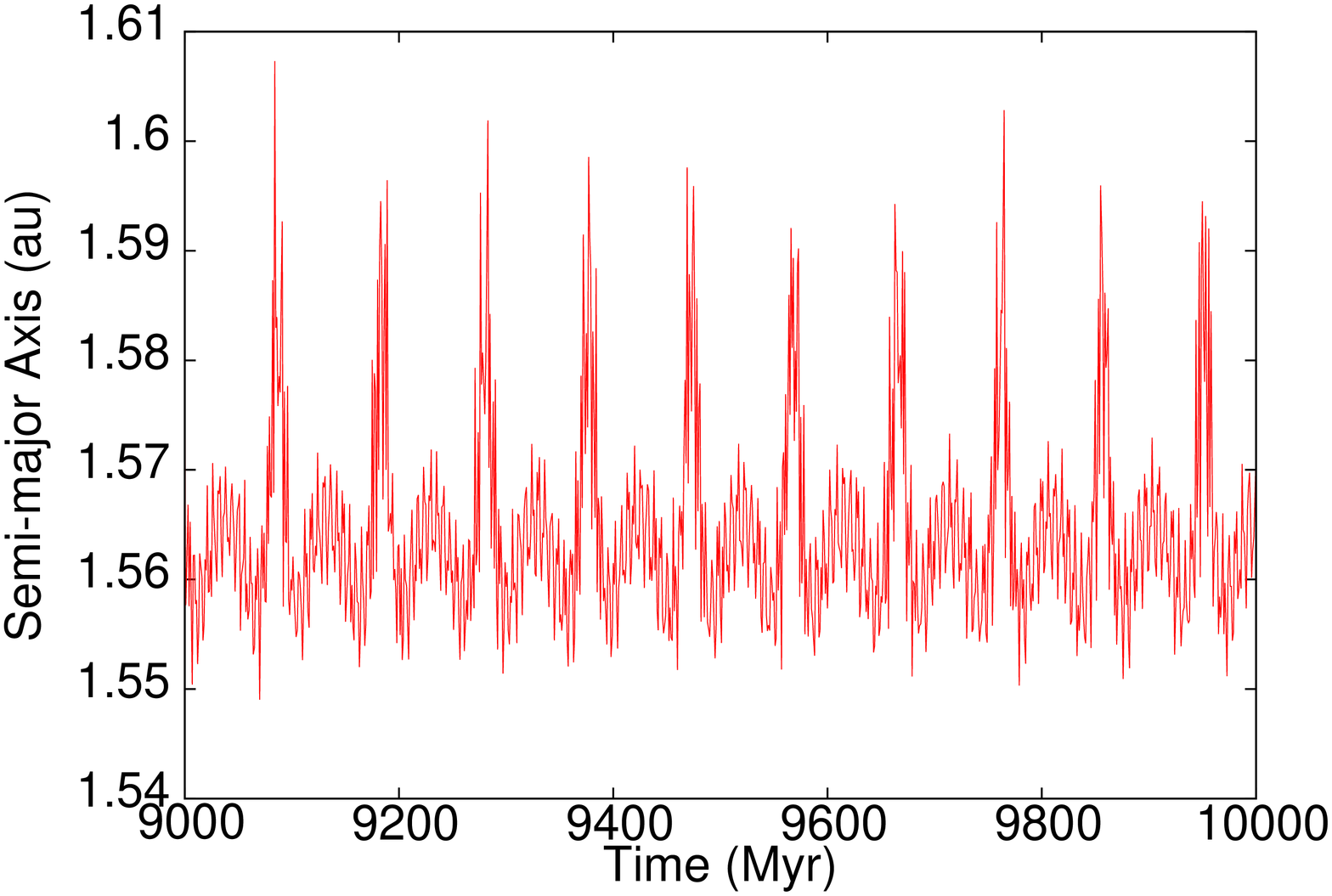,angle=270,width=8cm}
\epsfig{file=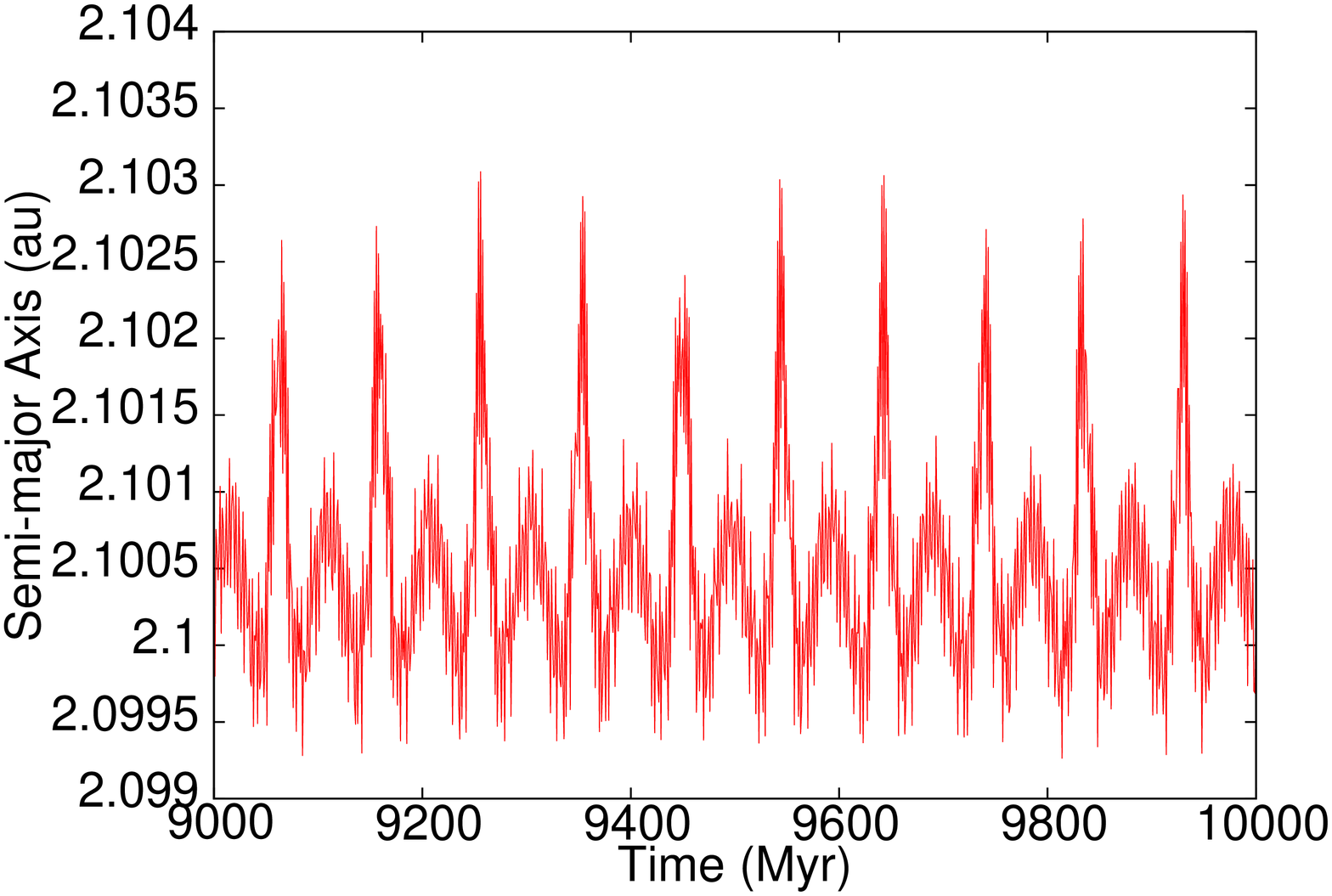,angle=270,width=8cm}
\end{tabular}
\caption{
Time evolution of the semi-major axis for the `Hilda'-planet (upper panel, left) and
the gas giant (upper panel, right) for the first hundreds of millions of years.
Results for the time interval between 9 to 10 Gyr for the `Hilda'-planet
(lower panel, left) and the gas giant (lower panel, right) are given as well. 
}
\label{Fig9}
\end{figure*}

Finally, we started a fictitious `Hilda'-planet in the second stable window,
which is again within the system's HZ (see Table~\ref{table1}).
The integration time was set to 10~Myr and the orbital elements of all
planets were checked regarding stability with special emphasis on the
`Hilda'-planet and the nearby gas giant, 47~UMa~b.  Figure~\ref{Fig9}
depicts the orbital evolution for the `Hilda'-planet during the first Myr
(left upper plot).  There is a main period visible, which is $P \sim 70$~kyr
with two different amplitudes for the semi-major axis.  After 9~Myr
a slight change in the periods occurs; for the gas giant the signal
is akin to a diminishing period with time (right plots).

\begin{figure*} 
\centering
\begin{tabular}{c}
\epsfig{file=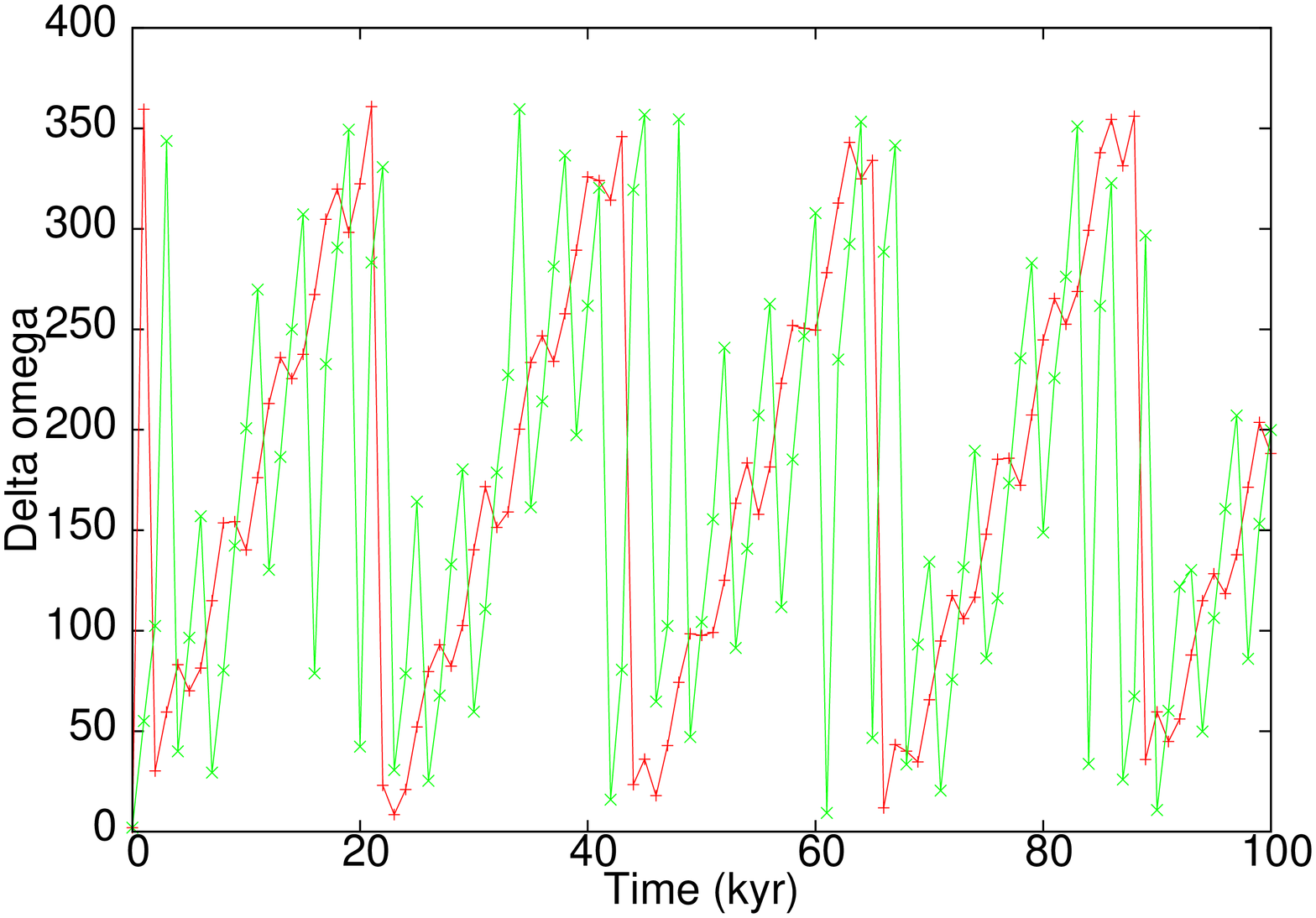,angle=270,width=8cm} \\
\epsfig{file=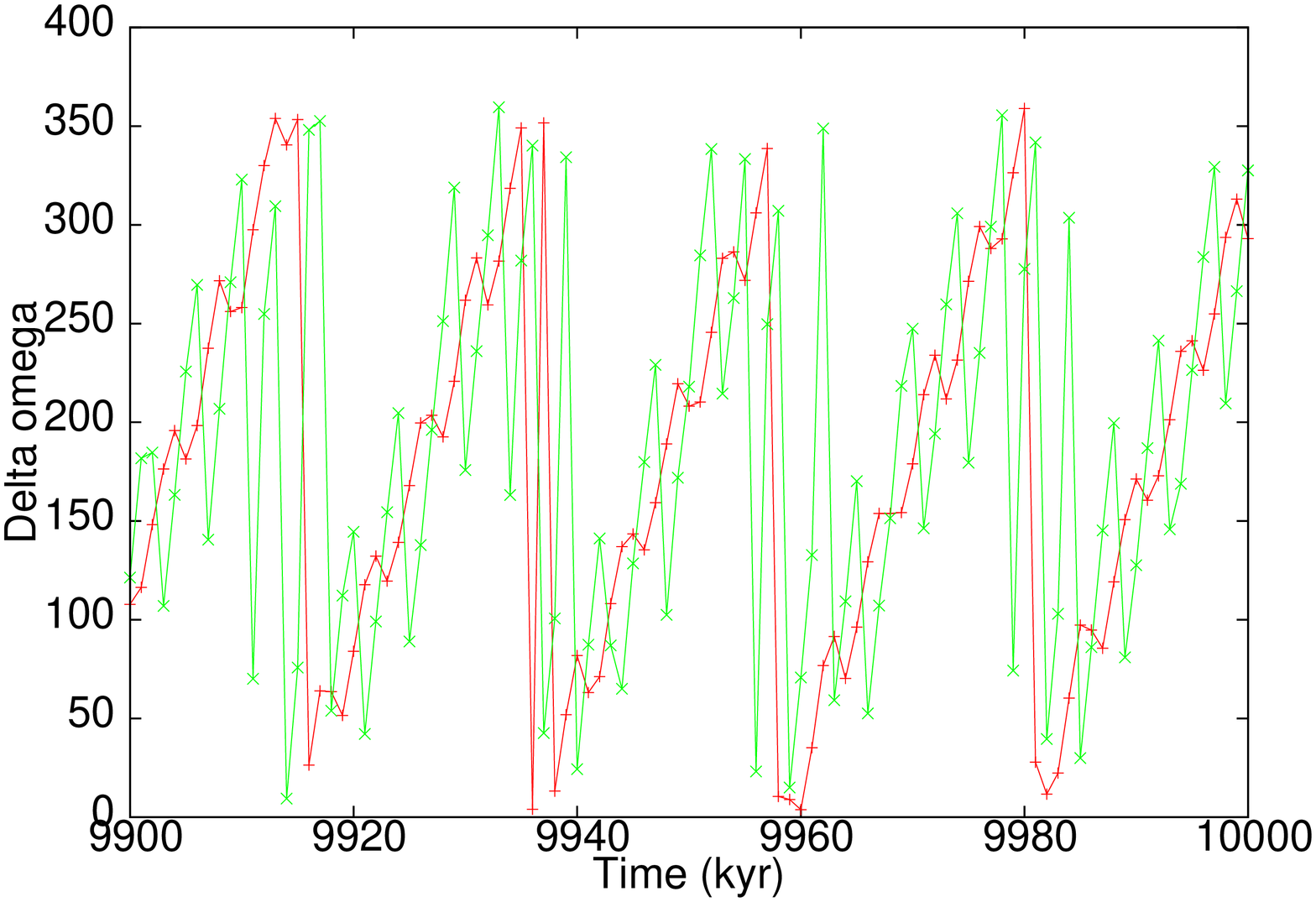,angle=270,width=8cm}
\end{tabular}
\caption{
Comparison of the time evolution of the difference
$\tilde{\omega}_{\rm planet} - \tilde{\omega}_{\rm Hilda}$ for the first 100~kyr and
last 100~kyr during the 10~Myr integration.  $\Delta \omega$ is depicted as a function of time.
The red line and green line indicate the gas giant and the `Hilda'-planet, respectively.
}
\label{Fig10}
\end{figure*}

In Figure~\ref{Fig10} we plotted the difference
$\tilde{\omega}_{\rm planet} - \tilde{\omega}_{\rm Hilda}$; it indicates
how the orbit of the fictitious `Hilda'-planet is connected to one
of the system planets, i.e., 47~UMa~c. The stability of the `Hilda'-planet
is caused by the coupling of this planet at 1.58~au to the Jupiter-type
planet at 2.1~au.  The reason for the stability is due to the
secular apsidal resonance (1:1, with a period of 20~kyr) also observed for
other planetary systems as described in earlier work \cite[e.g.,][]{goz03,ji03}.


\section{Summary and Conclusions}
\label{sec:summary}

The focus of this study is to explore the orbital dynamics of possible
exocomets in the system of 47~UMa.  This star exhibits various similarities
to the Sun, as evidenced by its spectral type, effective temperature,
metallicity, and mass --- although it has been determined that 47~UMa is
more evolved compared to the Sun while still being on the main-sequence;
see Sect.~\ref{sec:system} for references and details.  Furthermore, 47~UMa
is host of three Jupiter-mass planets, discovered between 1996 and 2010.
In addition, simulations have indicated the possibility of Earth-mass planets
in the system \citep[e.g.,][]{the02}, although there are no discoveries yet.
These putative terrestrial planets are at semi-major axes $a \lta 1.5$~au,
with the range of those results somewhat depending on the adopted model
parameters.  In our work, we consider hypothetical terrestrial planets
with initial distances of 1~au, 1.25~au, and 1.584~au; the later object is
also referred to as `Hilda'-planet.   Our work is motivated by previous
acclaimed detections of comets in various extrasolar systems, as reported
by, e.g., \cite{kie14}, \cite{coo16}, \cite{rap18}, among others.

Our sets of integrations yielded the following results:

\noindent
(1)  The general behavior of cometary collisions regarding the hypothetical
terrestrial planets is determined by 47~UMa~b, the closest in and most massive
of three Jupiter-type 47~UMa planets.  This planet is significantly closer to
47~UMa's HZ (compared to Jupiter in the Solar System), and thus a larger impact
on the orbital stability of HZ Earth-mass planets. 

\noindent
(2) As expected, the probability of a comet-planet collision was found to
depend on the cometary initial conditions (i.e., inclination and eccentricity).
Comets with relatively low initial eccentricities were unable to reach the inner
planetary system; thus, there were unable to collide with any of the assumed
Earth-mass planets unless they were scattered inward by the 47~UMa system planets.

\noindent
(3) Regarding the Earth-mass planet at 1.584 au (i.e., the `Hilda'-planet),
no collisions are found to occur at all, which makes water transport to Hilda
analogs through cometary collisions in a system akin to 47~UMa impossible.

\noindent
(4) Motivated by the structure of the Solar System, we also checked if the
comets form families; however, no families were found.  But the system might
contain different comet populations.  According to our simulations, there is
a small population of comets being scattered to orbits with very low eccentricities
and small semi-major axes, which might be analogous to the Jupiter-family comets.

\noindent
(5) Moreover, it was found that comets can be captured in close orbits, implying
that they may disintegrate as part of their future dynamic evolution.

Thus, the overall picture implies that --- following the premise that
all the water on the planets arises from cometary impacts ---  Earth-mass
planets in the 47~UMa system are expected to constitute ``land worlds".
Those planets would still be potentially habitable.
Land worlds, though from a global
perspective considered unlikely, as argued by \cite{sim17}, are expected
to possess a range of well-pronounced atmospheric and geodynamic features.
For example, \cite{abe11} studied those planets (also referred to as
``desert worlds") via three-dimensional global climate models and
concluded that they imply wider HZs compared to Earth-sized planets for
a given star.  Based on this work, Venus --- at a distance of 0.723~au
(semi-major axis), which would translate to 0.88~au for the 47~UMa system
--- might have been habitable as recently as 1~billion years ago.

In a separate approach, \cite{ver17} pointed out that studies about the
history of terrestrial biological evolution showed that significant
evolutionary innovation occurred as a consequence of land colonization,
which is Earth-based evidence for the significance of land words in
biological contexts.   As indicated in this work, 11 of 13 major
post-Ordovician innovations appeared first or only on land.  Although
the evolutionary patterns of life on exoplanets (if present) are expected
to be fundamentally different from terrestrial life forms, land worlds
still deserve further attention.

On the other hand, terrestrial planets located in 47 UMa's HZ might
still be able to harbor considerable amounts of water (i.e., on the
surface or subsurface), despite the relative inefficiency of exocomets
if other processes for the origin of water would have occurred.
For example, \cite{ray17} in their study about the Solar System argued
that water in the inner part of the system could have been delivered
as a by-product of the giants' formation process.  They found that
as a gas giant's mass increases, the orbits of nearby planetesimals
(supposedly, many of them with considerable water contents) are
destabilized and gravitationally scattered in all directions,
including the domain of terrestrial planets in the center star's
vicinity.

Therefore, regarding 47~UMa, even if the amount of water
proliferated to Earth-mass planets via exocomets is
small (or zero, or close to zero, as for the `Hilda'-planet), there
are still prospects for water-based habitability.  In fact, some previous
studies for the Solar System comets of, e.g., Halley, Hyakutake, Hale-Bopp,
and 67P/Churyumov-Gerasimenko indicate that Earth's water originating
solely from comets as implausible, considering their isotope ratios of
deuterium to protium (D/H ratio) measurements; see \cite{ebe95},
\cite{boc98}, \cite{mei99}, and \cite{alt15}, respectively.  A
recent review by \cite{ale18} conveys D/H values for a total of
13 Solar System comets, encompassing both Oort Cloud and Jupiter Family
comets.  Typically, the D/H ratios were found to be a factor of 1.5 to 3
times higher than the terrestrial value, which is $1.49 \pm 0.03 \times 10^{-4}$
\citep{lec98}.  However, exceptions exist.  For example, the D/H ratio of
C/2014~Q2 (Lovejoy) perfectly agrees with that of Earth \citep{biv16}.

Thus, it is both timely and appropriate to continue developing
different scenarios for the proliferation of water to possible
terrestrial planets in systems such as 47~UMa, which are
characterized by the existence of giant planets as well as
the possibility of terrestrial planets in the systems' HZs.


\acknowledgments
This research is supported by the Austrian Science Fund (FWF) through
grant S11603-N16 (B. L. and R. D.).  Moreover, M. C. acknowledges support
by the University of Texas at Arlington.  Additionally, we wish to thank
the anonymous referee for their helpful comments.


\clearpage



\clearpage

\clearpage

\clearpage

\clearpage

\clearpage

\clearpage

\clearpage

\clearpage

\clearpage

\clearpage

\clearpage


\begin{deluxetable}{lcc}
\tablecaption{Stellar Parameters}
\tablewidth{0pt}
\tablehead{
Parameter & Unit & Value
}
\startdata
 Spectral Type      & ...           &   G1~V           \\         
 $T_{\rm eff}$      & (K)           &   5890           \\     
 $M_{\ast}$         & ($M_\odot$)   &   1.03           \\
 $R_{\ast}$         & ($R_\odot$)   &   1.17           \\
 Metallicity        & (\% solar)    &   110            \\
 Age                & (Gyr)         &   6.03           \\
 $d_{\rm in}$~(HZ)  & (au)          &   0.91           \\
 $d_{\rm out}$~(HZ) & (au)          &   2.13           \\
\enddata
\tablecomments{
All variables have their usual meaning.  Here $d_{\rm in}$ and $d_{\rm out}$
indicate the limits of the optimistic HZ, also referred to as RVEM limits;
see \cite{kop13}.  However, the outer part of the HZ is unavailable for
hosting planets as they would be orbitally unstable due to the gravitational
influence of 47~UMa~b.
}
\label{table1}
\end{deluxetable}


\clearpage

\begin{deluxetable}{lccccl}
\tablecaption{47~UMa Planets}
\tablewidth{0pt}
\tablehead{
Name & Alternate Name & $a$ & $e$ & $m$ & Discovery \\
\noalign{\smallskip}
\hline
\noalign{\smallskip}
... & ... & (au) &  ...  &  ($M_J$) & ...
}
\startdata
47~UMa~b & Taphao Thong &  2.1  & 0.032  &   2.53 & \cite{but96}  \\
47~UMa~c & Taphao Kaew  &  3.6  & 0.098  &   0.54 & \cite{fis02}  \\
47~UMa~d & ...          & 11.6  & 0.16   &   1.64 & \cite{gre10}  \\
\enddata
\label{table2}
\end{deluxetable}


\clearpage

\begin{deluxetable}{lccc}
\tablecaption{47~UMa Planet Data}
\tablewidth{0pt}
\tablehead{
Name & $R_{\rm Hill}$ & $R_{\rm Hill}$ & $r$ \\
\noalign{\smallskip}
\hline
\noalign{\smallskip}
 ... & (au) &  ($10^6$ km)  &  ($10^3$ km)
}
\startdata
47~UMa~b &  0.195  & 29.23  &   93.9  \\
47~UMa~c &  0.200  & 29.94  &   56.1  \\
47~UMa~d &  0.934  & 139.7  &   81.2  \\
\enddata
\tablecomments{
Hill radii and planetary radii for the known planets of 47~UMa.
Intermediate values are given based on the densities
$\rho_1$ = 1 $\mathrm{g~cm}^{-3}$ and $\rho_2$ = 2
$\mathrm{g~cm}^{-3}$.
}
\label{table3}
\end{deluxetable}


\clearpage

\begin{deluxetable}{lccc}
\tablecaption{Parameter Ranges for the 47~UMa Comet Integrations}
\tablewidth{0pt}
\tablehead{
Quantity & Lower Bound & Upper Bound & Increment
}
\startdata
Semi-major axis $a$~(au)       &   80      &  200     &  10      \\
Eccentricity    $e$            &    0.915  &    0.990 &   0.005  \\
Inclination     $i$~($^\circ$) &    0      &   180    &  10      \\
\enddata
\label{table4}
\end{deluxetable}


\clearpage

\begin{deluxetable}{lccccc}
\tablecaption{Water Transport}
\tablewidth{0pt}
\tablehead{
... & nr (comets) & nr (coll) & \%EOs (1~Myr) & \%EOs (6~Gyr) & GLs (6~Gyr)
}
\startdata
  1 au    &  2 165 972 & 12 & 0.000016 & 0.093 & 61.7   \\
  1.25 au &  3 368 691 &  9 & 0.000012 & 0.073 & 48.5   \\
`Hilda'   & 68 429 091 &  0 & 0        & 0     & 0      \\
\enddata
\tablecomments{
Overview on the water transport for the different integrations.
Numbers are about comets for the full range of inclinations and an
initial semi-major axis of 80~au.  The amount of water transported is given
for a timespan of 1~Myr as well as for the 47~UMa system's lifetime of 6~Gyr
through extrapolation.  Results are expressed relative to Earth's Oceans
(\%EOs) and the Great Lakes (GLs).
}
\label{table5}
\end{deluxetable}


\clearpage

\begin{deluxetable}{lccccc}
\tablecaption{Water Transport, cont'd}
\tablewidth{0pt}
\tablehead{
... & nr (comets) & nr (coll) & \%EOs (1~Myr) & \%EOs (6~Gyr) & GLs (6~Gyr)
}
\startdata
  1 au    &   2 362 287 & 13 & 0.000017 & 0.107 & 70.5   \\
  1.25 au &   3 524 063 & 12 & 0.000016 & 0.093 & 61.7   \\
 `Hilda'  & 189 883 894 &  0 & 0        & 0     &  0     \\
\enddata
\tablecomments{
Same as in Table \ref{table5}, but for all initial semi-major axes
in the assumed range (see Table~\ref{table4} and
Sect.~\ref{sec:orbitaldistribution}).  The results show similar trends
as before.
}
\label{table6}
\end{deluxetable}

\end{document}